\documentclass[12pt,preprint]{aastex}
\shorttitle{Late type active stars: FR Cnc, HD 95559 and LO Pegasi}
\shortauthors{Pandey et al.}
\begin{document}
\title{Optical and X-ray studies of chromospherically active stars 
: FR Cancri, HD 95559 and LO Pegasi}
 
\author{J. C. Pandey\altaffilmark{1}, K. P. Singh\altaffilmark{2}, S. A.
Drake\altaffilmark{3} and  R. Sagar\altaffilmark{1}}

\altaffiltext{1}{Aryabhatta Research Institute Of Observational Sciences (ARIES)
, Naini Tal-263 129, India; jeewan@aries.ernet.in, sagar@upso.ernet.in}
\altaffiltext{2}{Tata Institute of Fundamental Research, Mumbai - 400 005, India; singh@tifr.res.in}
\altaffiltext{3}{USRA \& Code 662, NASA/Goddard Space Flight Center, Greenbelt,
MD 20771, USA; drake@olegacy.gsfc.nasa.gov}

\date{Received ...../ accepted .....}

\begin{abstract}
We present a multiwavelength study of three chromospherically active stars, 
namely FR Cnc (= BD +16\degr 1753), HD 95559 and 
LO Peg (=BD +22\degr 4409), including newly obtained optical photometry,
 low-resolution optical spectroscopy for FR Cnc, as well as archival 
IR and X-ray observations.  

The BVR photometry carried out 
during the years 2001 - 2004 has found significant photometric variability 
to be present in all three stars. For FR Cnc, a photometric
period $0.8267 \pm 0.0004 \rm~{d}$ has been established.  The strong  
variation in the phase and amplitude of the FR Cnc light curves when folded on
this period implies the presence of evolving and migrating spots or spot groups 
on its surface. Two independent spots with migration periods 
of 0.97 and  0.93 years respectively are inferred. 
The photometry of HD 95559 suggests the formation of a spot (group) 
during the interval of our observations.
We infer the existence of two independent spots or groups in
the photosphere of LO Peg, one of which has a migration period of 1.12 years.

The optical spectroscopy of FR Cnc carried out during 2002-2003, reveals
the presence of strong and variable Ca II H and K, H$_\beta$ and $H_{\alpha}$ 
emission features indicative of high level of chromospheric activity.
The value of 5.3 for the ratio of the excess emission in $H_{\alpha}$ to
$H_{\beta}$, $E_{H_{\alpha}}/E_{H_{\beta}}$, suggests
that the chromospheric emission may arise from an extended off-limb region. 

We have searched for the presence of color excesses in the near-IR 
JHK bands of these stars using 2MASS data, but none of them appear to have 
any significant color excess. 
We have also analyzed archival X-ray observations of HD 95559 and LO Peg 
carried out by with the ROSAT observatory. 
The best fit models to their X-ray spectra imply the 
presence of  two coronal plasma components of differing temperatures and
with sub-solar metal abundances.  The inferred emission measures and 
temperatures of these systems are similar to those found for other 
active dwarf stars.

The kinematics of FR Cnc suggest that it is a very 
young (35 - 55 Myrs) main-sequence star and a possible member of the IC 2391
supercluster. LO Peg also has young 
disk-type kinematics and has been previously suggested to be a member
of the 100 Mys old Local Association (Pleiades Moving Group).
The kinematics of HD 95559 indicate it is a possible 
member of the 600 Myrs old Hyades supercluster. 
\end{abstract}

\keywords{X-ray: stars --- stars: activity --- binaries: general --- stars: individual (FR Cnc, LO Peg, HD 95559)}

\section{Introduction}
 Differential rotation in late-type stars having convective envelopes
drives a magnetic dynamo leading to strong chormospheric emission and the
formation of a corona.  The Solar corona, though readily visible because
of its proximity, is only about $10^{-6}$ of the Solar photospheric emission.
Rapidly rotating late-type stars, however, display an extremely enhanced
coronal activity when compared with that of the Sun. 
Strong X-ray and nonthermal radio emission in late-type stars are well-known  
indicators of enhanced coronal activity (Drake, Simon \& Linsky  1992). Ayres and 
collaborators have shown
(e.g., Ayres et al. 1995) that the emission from coronae ($T_e \sim 10^{6-7}$K) and
chromospheres ($T_e \sim 10^4$K) are closely correlated, and thus stars with 
intense coronae will also have strong chromospheres. While many active stars 
have been identified as such through their above-average X-ray and radio 
emission, it is only through detailed optical photometric and 
spectroscopic studies that their activity can be classified into known types, 
such as the RS CVn, BY Dra and FK Comae classes. The common characteristic of
all these various classes of active star, be they single stars or binaries, 
is rapid rotation: single, rapidly rotating stars are either young stars
which have not yet lost most of their angular momentum or (in a few rare
cases) are the results of the merger of a close binary system, while rapid
rotation in close binaries is the natural result of spin-orbit tidal coupling
and can occur even in middle-aged and/or old stars.

RS CVn types are binary systems 
that contain a hotter component with spectral type F- to  G- and 
luminosity class V or IV, and a cooler component that is
usually a sub-giant or a giant K-type star (Hall 1976). RS CVns have
been further subdivided into three groups according to their orbital 
period (P) : short period RS CVn (P $\leq$ 1 d); {\it classical} 
RS CVn (1d $<$ P $\leq$ 14d); and long period systems usually 
containing at least one cool giant component (P $>$ 14d). 
BY Dra type stars have properties similar to the RS CVn systems, 
but consist of a late dKe or dMe type star with orbital or rotational 
period in the range of $\approx 0.5$ to $\approx 20 \rm~{d}$. 
As originally defined by Bopp \& Fekel (1977), BY Dra types may 
include active {\bf single} main-sequence stars as well as members of 
detached binary systems. (In the alternate definition of active binary systems 
of Fekel, Moffett \& Henry 1986, there is no restriction on the spectral type of the 
companion star or on the orbital period for RS CVn binaries, except that the
more active star is in an evolved evolutionary state, and systems
with F- and G type dwarf primaries are now classed as BY Dra binaries
rather than as a sub-type of RS CVns).  FK Com stars are a rather rare
class of apparently single, rapidly rotating late-type (G or K) giant stars 
which were first described by Bopp \& Stencel (1981): the origin of these 
stars is still poorly understood, with the hypothesis that they are the
remnants of coalesced binaries being the most widely accepted. 
As a group, the late spectral type BY Dra stars tend to more often exhibit
H$_{\alpha}$ emission and more frequent flaring activity 
than the RS CVn systems, but this is probably mostly due to the greater 
contrast of their chromospheric emission lines relative to their weaker 
photospheric emission.
Both RS CVns and BY Dra binaries typically rotate synchronously with their
orbital period, but there  are $40+$ known systems (Glebocki \& Stawikowski 
1997) which are known to rotate asynchronously, e.g., the prototype K3.5V $+$
K 3.5V binary system BY Dra which has a rotational period $P_{rot}$ of 3.84
and an orbital period of $5.98 \rm~{d}$.

The soft (0.1 - 2.0 keV) X-ray luminosities ($L_{x}$) of RS CVn systems  
are generally in the range of $10^{29} - 10^{32}$ erg s$^{-1}$ (Drake, Simon \& Linsky
1989), while BY Dra stars tend to have $L_{x} \approx 10^{29} - 10^{30}$ 
erg s$^{-1}$ 
(Dempsey et al. 1993). In either case the value of $L_{x}$ for 
these stars are $ \approx 10^{2-5}$ times the X-ray luminosity of 
the Quiet Sun. The radio luminosity ($L_{6cm}$) of RS CVn systems
ranges from $10^{14}$ to $10^{17}$ erg s$^{-1}$ Hz$^{-1}$ 
(Morris \& Mutel 1988; Drake, Simon \& Linsky 1989), while for the BY Dra types it is 
typically $10^{12-15}$ erg s$^{-1}$ Hz$^{-1}$ (Caillault, Drake \& Florkowski 1988; G\"{u}del et 
al. 1993), much more than the Quiet Sun value of $3 \times 10^{10}$ 
erg s$^{-1}$ Hz$^{-1}$ in both cases. In this paper we present a detailed
investigation based on our extensive optical observations and archival
X-ray data of three chromospherically active stars selected on the basis
of their strong X-ray and radio fluxes. These stars are FR Cnc, HD 95559 
and LO Peg.

The star BD +16\degr 1753 (= MCC 527) first came to be noticed 
as a potential active star when
it was identified as the likely optical counterpart of a soft X-ray source in 
the Einstein Slew Survey, 1ES 0829+15.9, with an observed X-ray flux of $\approx
10^{-11}$ erg s$^{-1}$ cm$^{-2}$ (Elvis et al. 
1992; Schachter et al. 1996). This X-ray source was rediscovered in the ROSAT
All-Sky Survey (RASS) at a somewhat weaker (but more statistically 
significant) level of $2 \times
10^{-12}$ erg s$^{-1}$ cm$^{-2}$, and dubbed 1RXS J083230.9+154940 in the RASS
Bright Source Catalogue (Voges et al. 1999).
In the photometric notes annex of Hipparcos catalogue (Perryman et al. 1997), 
it was regarded as an unsolved variable star and given the name FR Cnc. 
It was recently classified as a BY Draconis type in the 74th special 
Name-List of the variable stars (Kazarovets et al. 1999). 
The implied X-ray luminosity of $2 - 12 \times 10^{29}$ erg s$^{-1}$ and ratio 
of X-ray to bolometric luminosity $f_{x}/f_{bol}$ of $ \geq 10^{-3.3}$ both
show that FR Cnc has an active corona at or near the saturation limit 
of $f_{x}/f_{bol}$ of $ \approx 10^{-3}$ (Schachter et al. 1996).
Recently, Upgren, Sperauskas \& Boyle (2002) have given two measurements of the 
radial velocity of FR Cnc (see Table 5) which differ only by an amount
of the order of the measurement error, and the authors thus conclude that it
is not a binary system.  A preliminary report on FR Cnc based on a limited 
subset of the data was presented by Pandey et al. (2002).

HD 95559 (= BD 23\degr 2287) has recently been shown to be a double lined
spectroscopic binary with an orbital period of $1.5260 \rm~{d}$. 
With a photometric period of $1.5264 \rm~{d}$ it is tightly synchronized 
to the orbital motion (Fekel \& Henry 2000). Previous reports that
this system has a $2.9 \rm~{d}$ photometric period (Jeffries, Bertram \& Spurgeon 
1994; Strassmeier et al. 2000) appear to have been the detection of 
the $1 \rm~{d}$ alias of the $1.526 \rm~{d}$ orbital period. ~HD 95559 is a pair of K1 V
stars with a Li-based age younger than the Hyades cluster. 
Fekel \& Henry (2000) conclude that optical variability in this 
system is due to the rotational modulation of the star spots and thus identify
it as belonging to the (binary) BY Dra type.

LO Peg (=BD+22\degr 4409) is a single young K5V - K7V  type star and is a member of 
the local association (Jeffries \& Jewell 1993; Montes et al. 2001). It is an 
active star showing strong $H_{\alpha}$ and Ca II H and K emission lines
(Jeffries et al. 1994). ~Jeffries et al. (1994) proposed six probable 
rotational periods, and  stated that the periods of 0.3841 d and 0.42375 d are more
likely. Subsequently, Robb \& Cardinal (1995) eliminated completely the possibility of the 
0.38417 d period.  Evidence of an intense downflow of material and 
optical flaring on LO Peg have been presented by Eibe et al. (1999).
Recently, Zuckerman, Song \& Bessel (2004) has identified LO Peg as a member of a group of $\sim 50$ Myr
old stars that partially surround the Sun.

The paper is organized as follows: in  \S 2, we describe the primary 
observational datasets that we have analyzed, and the methods of data 
reduction. The  optical light curve and the period analysis of 
the star FR Cnc are discussed in \S 3.  In \S 4, we present the folded light
and color curves of all three stars and provide an interpretation 
for the  drift in phase of their wave minima. In \S 5, we discuss 
the low-resolution optical spectra of FR Cnc that we have obtained.  
The ROSAT X-ray spectra of the stars HD 95559 and LO Peg  are discussed 
in \S 6, and \S 7 contains a comparison of their X-ray properties with  
other similar stars. The  physical parameters, JHK photometry and kinematics
of the stars FR Cnc, HD 95559 and LO Peg are described \S 8, 9, and 10,
respectively, while we present our conclusions in \S 11.    

\section{Observations and Data reductions}
\subsection{Optical Photometry}

FR Cnc, HD 95559 and LO Peg  were  observed in Johnson B, V and Cousins R filters 
from  2001 to  2004 at the State Observatory of Naini Tal 
(now renamed as Aryabhatta Research Institute of Observational Sciences, 
ARIES).  The observations were made with the 104-cm Sampurnanand 
telescope (SNT) to which a $2k \times 2k$ CCD  
(field of view $\sim 13 \arcmin \times 13 \arcmin $) in the years 2001 -2003,
and a $1k \times 1k$ CCD  (field of view $\sim 6\arcmin \times 6\arcmin $) 
during the season 2003-2004, were  attached. A log of the observations
is given in Table 1.
A number of CCD frames were taken on each night with different 
exposure times ranging from 10 to 120 secs depending upon the seeing 
conditions and the filter used.  Several bias and twilight flat frames 
were taken during the observing runs.
Bias subtraction, flat-fielding and aperture photometry were performed using 
IRAF\footnote{IRAF is distributed by National Optical Astronomy 
Observatories, USA}. 
For each star,  differential photometry in the sense of variable minus the comparison 
was done as all the program, comparison and check stars were in the 
same CCD frame. 

 UBVRI observations of FR Cnc along with that of Landolt (1992) 
standard SA 98 region were obtained on 22 February 2004 for photometric 
calibration. The average magnitudes of FR Cnc in the UBVRI filters 
were $12.19 \pm 0.02, 11.34 \pm 0.01, 10.24 \pm 0.01, 9.567 \pm 0.009,$ 
and $8.89 \pm 0.006$, respectively. 
The present photometry is well  matched to the Weis (1993) photometry.
LO Peg was also observed for photometric calibration using PG 2213-006
standard fields (Landolt 1992) on 03 October 2002 in B, V, R and I band.
The corresponding average magnitudes are $9.86 \pm 0.01$, $8.82 \pm 0.01$, $8.22 \pm 0.01$ and $7.78 \pm 0.01$, respectively.

\subsection{Optical Spectroscopy}
Spectroscopic observations were carried out on 2002 November 15 and 2003 
January 20 to 24 at the Vainu Bappu Observatory, Kavalur with the OMR 
spectrograph fed by the 234-cm Vainu Bappu Telescope (VBT). 
The data were acquired with a $1 k\times 1 k$ CCD camera 
of $ 24 ~\mu m$ square pixel size, covering a range of $\sim 1200 ~\AA$ and 
having a
dispersion of $1.25 ~\AA/pixel$.  A signal-to-noise ratio 
between 20 to 30 was achieved. During these observations we took
ten spectra of FR Cnc in the $H_\alpha$ region, six in the $H_{\beta}$ region
and five in the Ca II H and K region. A total of 21 spectra were thus taken.
HD 26795, a K3V type star, was observed as the standard. 

The spectra have been extracted using the standard reduction procedures in the
IRAF packages (bias subtraction, flat fielding, extraction of the spectrum
and wavelength  calibration using arc lamps). The spectral resolution
was determined by using emission lines of arc lamps taken on the same nights. 
Spectral resolution ($\delta \lambda$) of $2.7 ~\AA$ at $6300 ~\AA$ and 
$3.7 ~\AA$ at $4000 ~\AA$ was achieved. 
All the spectra were normalized to the continuum and equivalent widths (EWs) 
for emission lines were computed using IRAF task splot. 
 The error in the measurements of EWs of the $H_\alpha$, $H_\beta$ and
CaII H and K lines was computed by measuring the EWs of 4 moderate absorption
features at 6494, 6163, 6122 and 6103 \AA~ in each spectra. The standard 
deviation for each absorption feature was computed from the 10 spectra
available. The mean of the standard deviations thus obtained was 0.04 \AA,
and was taken as the error in the measurement of the EWs of all the absorption
as well as the emission line features (see also Padmakar et al. 2000).

\subsection{X-ray}

The stars HD 95559 and LO Peg were observed serendipitously in pointed
observations of the ROSAT PSPC 
detector. This instrument had an energy range from 0.1-2.4 keV with a
rather low spectral resolution ($\Delta E/E \approx 0.42$ at 1 keV).
A full description of the X-ray telescope and detectors can be
found in Tr\"{u}mper (1983) and in Pfeffermann et al. (1987).
The ROSAT X-ray data for these two stars were obtained from the 
High Energy Astrophysics Science Archive Research Center (HEASARC).
The ROSAT observational parameters are given in Table 1. 
LO Peg was observed on two occasions, the first on 1993 November 11-13, and 
the second a month later on 
1993 December 7-8, while the HD 95559 observation consisted of two short 
exposures in 1993 June separated by a week. (FR Cnc was not observed in any of the 
pointings made by ROSAT, but it was, however, detected in the ROSAT All-Sky 
Survey phase, as has already been mentioned in the introduction).
HD 95559 and LO Peg were offset from the  center of the PSPC field
by 38.4 and 0.31 arcmin, respectively.
Spectra of both the stars were accumulated from on-source counts obtained from 
circular region on the sky centered on the X-ray peak. The background was 
accumulated from several neighboring regions at nearly the same offset
as the source.

\section{Light curves and period analysis of FR Cnc}

\subsection{Present light curve}

We obtained nine light curves, with each light curve corresponding to an 
observational run that was nearly continuous.  
Figure 1 shows the light curves of FR Cnc for each epoch.
The mean epoch of the light curves, the observed maximum 
($\Delta V_{max}$) and minimum ($\Delta V_{min}$) 
in the V band, peak to peak amplitude 
($\Delta V = \Delta V_{max} - \Delta V_{min} $), and phase minima 
($\theta_{min}$) are listed in Table 2.   $\theta_{min}$ was determined 
by a linear least square fitting of the second order polynomial at the minimum
of each light curve.  It appears that the shape and the amplitude of 
the light curves are changing during the observing season. 

During the observing season 2001-2003
for the star FR Cnc the comparison and check stars were TYC 1392 2110 1 (= S1) and 
 USNO-A2.0 1050-05766589 (= C), respectively. 
BD +16\degr 1751 (= S2)  of the spectral type G0 was taken as the comparison star 
during the observing season 2003-2004. The mean difference between the 
(S2-C) and (S1-C) in each BVR filter was added to each BVR data of 
2003-2004 season before plotting.
No significant light variation was detected between the different 
measures of comparison and check stars ($\Delta V_{c}$)
(see the bottom panels of 
Figure 3 (a to i)) indicating that the comparison stars are constant 
during the present observations. 
The nightly mean of the standard deviation ($\sigma$) of different measures 
of comparison and check stars in B, V and R filter was 0.007,0.007 and 0.008 mag,
respectively. Similarly, $\sigma$  determined for $\Delta(B-V)_{c}$ and  
$\Delta(V-R)_{c}$ was 0.006 and 0.01 mag.

Large gaps in the dataset lead to complications in the interpretation 
of the power density
spectrum as the true frequencies of the source are further modulated 
by the irregular and infrequent sampling defined by the window function 
of the data. Therefore, the calculation of the dirty power density spectrum 
from the light curve was followed by deconvolution of the window function 
from the data using one dimensional CLEAN (Roberts, Lehar \& Dreher  1987)
algorithm in Starlink's PERIOD package,  where we have assumed quasi-sinusoidal
light curve.  The power spectrum picks out the rotation 
period despite the noise and the fact that the light curve 
is not exactly sinusoidal (Bailer-Jones \& Mundt 2001).
The uncertainty in the period determination is set by the finite resolution 
of the power spectrum. 
This is determined as  $\rm{P}^{2}/(2t_{max})$, where P is period and 
$t_{max}$ is the duration of the observations (Roberts, Lehar \& Dreher 1987).
The CLEANed power spectra presented here were obtained 
after 100 iterations with a loop gain of 0.1.  Data for each epoch were analyzed
separately for the periodicity by using the CLEAN algorithm. 
Inset in each panel of  Fig. 1 shows the corresponding 
CLEANed power spectrum. The epoch and the period are mentioned 
inside each panel of Fig. 1. 
The period is found to be constant, within the uncertainty, for each epoch.

The entire data were also analyzed together using the CLEAN algorithm to
improve the period determination.  The top panel of Figure 2 shows the
discrete power spectra, with the corresponding window and CLEANed power spectra
shown in the middle and the bottom panel. 
The highest peak in the CLEANed power spectrum corresponds to a period 
of $ 0.8267 \pm 0.0004 \rm~{d}$ 

\subsection{Hipparcos light curve}
The light curve based on the Hipparcos photometric data 
is shown in Fig. 1 (j).  The total number of measurements with 
Hipparcos is 70 during three years (1990 March - 1992 November) 
of observations.
We have also analyzed the Hipparcos data using the same algorithm 
as used for our photometric data.
The CLEANed power spectrum is shown in the inset of  Fig. 1 (j). 
The highest peak here corresponds to a period of $0.815 \pm 0.002 \rm~{d}$,  
which is close to the period  determined from the present data.

\section{Folded Light and Color curves}

\subsection{FR Cnc}

The data were folded using the period of $0.8267\rm~{d}$, and an arbitrary 
epoch of JD 2451943.1980. Fig. 3 shows the plot of $\Delta V_{c}, 
\Delta B, \Delta V, \Delta R, \Delta (B-V)$ and $\Delta (V-R)$ as a function 
of phase at different epochs. The mean epoch is mentioned at the top
of each panel of Fig. 3.  As can be seen from Fig. 3, 
the phase coverage is fair to reasonably good in most of the light curves.    

The value of $\Delta V_{max}$ was nearly constant during most of the epochs
indicating that the brightness of the unspotted photosphere was constant from
epoch to epoch. During the epoch 'b' it went to its minimum value. At the same
time a smaller variation ($\Delta V = 0.063 \rm~{mag}$) during  this epoch is probably
associated with a continuous reduction of the spot coverage.  Light curve
during the epoch 'b' has two maxima whereas during the epoch 'c' it
has one flat maximum and one minimum indicating the formation of a new 
group of spots that further separate into two  groups of spots
during the epoch 'd' and 'e'. 
An increase in the amplitude by 0.05 mag during the epoch
'e' indicates an increase in the spot coverage on the surface of the star that
remained constant during epoch 'f'.  However, poor phase coverage during 
the epochs 'f', 'g' and 'i' does not allow us to follow this progress 
of spots properly. The activity level in our observation became very high 
during the epoch 'h', where the amplitude of variation became 0.21 mag and 
appeared to remain the same during the epoch 'i'.  
Variable size of the spot (or a group of spots) are indicated by the
variable depth of the light minimum in the light curves. 
The minimum of the light changed by  0.1 mag from epoch 'a' to epoch 'h'  
(see Table 2).

Comparing the light curves of FR Cnc from epoch 'a' to epoch 'i', 
a shift in the phase of the minimum and a variable amplitude is quite evident. 
Such cycle-to-cycle variation of both the amplitude and phase of the minimum 
has been seen for some rapidly rotating stars like LO Peg (Dal \& Tas 2003), 
Speedy Mic (Barnes et al. 2001),
AB Dor (Bos 1994; Donati \& Collier Cameron 1997) and PZ Tel (Innis et al. 1990).

 Variation in the color of the star is correlated with its magnitude 
i.e. the star becomes redder when fainter, and bluer when brighter, 
supporting the starspot hypothesis.  The significance of correlation 
has been calculated by determining the linear correlation coefficient, r,
between the magnitude and the colors. 
The value of r between V and (B-V), V and (V-R) and (B-V) and (V-R) was found
to be 0.11, 0.35 and 0.581 with corresponding probability
of no correlation being 0.00181, $1.51 \times 10^{-25}$ and 0, respectively.

\subsubsection{Phase of the light minimum}

The phase minimum of light ($\theta_{min}$) for the nine light curves in V band 
were plotted against the mean epoch listed in Table 2 and shown in left 
panel of Figure 4.  To illustrate the spot motions over several longitudinal
cycles, we define the longitude as longitude + n*360\degr  or ($\theta_{min}$ 
+ n*1), where n is the cycle number.
The phases of the light minima directly indicate the
mean longitude of the dominant groups of the spots.
The presence of two spots is clearly established by two well separated 
straight lines.  Both the spots are closer to each other during the 
epoch 'g', with a longitudinal separation of $\sim 83^{0}$. 
Both the spots are visible during all the epochs, except during the epoch 'c'. 
Poor phase coverage during the epoch 'f' could not establish the 
presence of the second spot. It is interesting to see the variable separation
of the two spots. This could be  due to the different latitudinal positions of 
the two spots and the presence of differential rotation in the star.
Due to differential rotation,
spots at different latitudes would give rise to a different migration 
rate for associated $\theta_{min}$ (Raveendran \& Mohin 1995).

The following relation was fitted to the data by the method of the least
squares:
\begin{equation}
\theta_{min} = \omega (t - t_{0}) + \theta_{0}
\end{equation}

where $\omega = 2 \pi/P$ is the  angular velocity of the spot or the 
rate of phase shift (degrees/day), P is the corresponding period in days,
t is time in days, $t_{0}$ is reference time, and $\theta_{0}$ is the 
reference spot longitude in degree.
Applying the above relation, the rate of phase shift for the spots A and B 
are determined as $1.02\pm 0.02$ degree/day and $1.06\pm 0.02$ degree/day, 
respectively. The corresponding migration period for the spot A and B 
are found to be $0.97 \pm 0.03$ years and $0.93 \pm 0.02$ years, respectively. 
 These migration periods, being so close to each other, indicate that the
spots that are located at different longitudes rotate with almost the same 
angular velocity. It thus appears that 
the observed phase minima of FR Cnc are well arranged in two permanent strips
with approximately the same slope (i.e. same migration period), 
which can be interpreted  as two long-lived active longitudes.

\subsection{HD 95559}

Stars HD 95467 and BD +33\degr 2294 were observed for the differential
photometry of the program star HD 95559.
 The data were folded  using the following ephemeris for HD 95559:

JD = 2450359.094 + 1.52599775E,

\noindent
where the epoch corresponds to a time when the more massive star is in front 
of the less massive star and period is spectroscopic orbital period (Fekel \&
Henry  2000). 
Variation in the B V R magnitude and colors of the program star
with respect to the comparison star are shown in Fig. 5a and 5b.
The nightly mean of $\sigma$ of the different measures of comparison
and check stars was determined 0.006 in each B and V filter,
and 0.007 in R filter. The comparison star was constant
during the observations as shown in the bottom panel of 
Figure 5a and 5b.
We have divided our observations into two epochs, namely,
 epoch 'a' ( 2001 February 3-5) and
epoch 'b' (2001 March 31 -  April 7).  The amplitude and the phase of the minima 
with the mean epoch of the observations of this star are shown in Table 2.

It is interesting to examine the light curves of HD 95559 during the observing 
season 2001. During the epoch 'a' the light curve has a flat topped maximum and 
only one minimum (Fig. 5a).
During the epoch 'b' that flat topped maximum developed
into a  minimum at a phase\ = 0.5. At the same time, the  amplitude 
of the variation increased by 0.025 mag. The position 
of the minimum during the epoch 'a' however, did not change from the 
phase 0.0. This directly indicates the formation of a new
group of spots at phase 0.5.
Light curves of HD 95559 during the observing season 1995-1997 had 
only one minimum and a sharp maximum (Fekel \& Henry 2000). 
 
It appears that the colors remain nearly constant during our observations
in HD 95559.  A larger amount of scatter is, however, seen during the epoch 
'b'. This may be due to the formation of the new group of spots on the surface
of the active star. The lack of brightness-color relation may be due to
the presence of bright facular or plages like regions accompanied by dark spots 
in any one or both components of this binary system.

\subsection{LO Peg}
\label{sec:LOPeg}

Photometric data of LO Peg were folded using the following ephemeris
as given by  Dal \& Tas (2003):

HJD =  $2448869.93 + 0.42375E$.

The folded light curves are shown in Figure 6. The observations separated by 
a year  were divided 
into two different epochs namely 'a' and 'b'.
The epoch 'a' has observations from 2001 September 28 to October 15, and 
epoch 'b' has the observations from 2002 September 28 to October 3.  The 
'b' epoch observations partly coincide with one set of observations
(2002 October 1-31) of Dal \& Tas (2003). 
Table 2 shows the parameters determined from the light curves of the LO Peg. 
Phase minimum computed from our photometry at epoch 'b'
differs by  0.26 from that of Dal \& Tas (2003) photometry. This may 
be due to the larger span of their observations, as phase mimimum changed 
significantly within a couple of rotational periods, probably as a result 
of change in the spot configuration.  As in the case of   FR Cnc 
no two light curves of LO Peg are similar in shape, size and amplitude,
which is seen quite commonly  in the BY Dra type of star.

A plot of the phase minima as a function of the mean epoch of the 
observations of LO Peg is shown in the right panel of  Fig. 4. 
The solid circles represent the observations made by Dal \& Tas (2003)
and the solid triangles represent our observations.
Two large groups of spots  are clearly seen.  We have fitted the eq. (1) by 
the linear 
least squares deviation method for the group of spots A and computed the
rate of phase shift. The open circle in the right panel of Fig. 4 was not used in the fitting.
The rate  of the phase shift for the
group of spots A was determined $0.85 \pm 0.03$ degree/day. The corresponding 
migration period is $1.12 \pm 0.05$ years.

From Fig. 6, it is clear that the color curves are strongly correlated with 
light curves, i.e. bluer at light maximum and redder at light minimum.
This indicate that the variability is due to the dark spots present on 
the surface of the star.

We have also checked whether the comparison star (BD +22\degr 4405) is variable or not.
Star USNO-B1.0 1133-0542608 was used as a check star. The 
$\sigma$ was computed between the different measures of comparison 
and check stars. Nightly mean of the $\sigma$ was  0.010, 0.009 and 0.007 in 
B, V, and R filter, respectively. Lower panels of Fig. 6a and 6b show the plot of
different measures of comparison and check stars and indicate that the 
comparison stars were indeed constant during the observations.  

\section{Chromospheric emission lines of FR Cnc}
 
In late type dwarfs, $H_{\alpha}$ emission is a good indicator 
of chromospheric activity. It has been suggested that the  detection 
of $H_{\alpha}$ emission above the continuum or even a filled-in emission 
is sufficient to indicate that K - M dwarf is a BY Dra variable (Bopp et al. 1981). 
Spectrum of FR Cnc in the $H_{\alpha}$ region is shown in Figure 7.
Spectrum of the star HD 26794 (V = 8.81 mag and Spectral type - K3V) is also  
shown in Fig. 7 for comparison. 
$H_{\alpha} $ is in emission at all phases in FR Cnc.  The measured equivalent width of the $H_{\alpha}$ 
emission feature and the corresponding JD and phase of FR Cnc
are listed in Table 3. The 
phase is determined using the photometric period of $0.8267 \rm~{d}$ and an 
epoch of JD 2451943.1980 for phase 0. The equivalent width is seen to vary 
from 1.2 to 1.8 $\AA$. 
Such a significant change in the $H_{\alpha}$ profiles and the equivalent 
width on a time scale as short as few hours has also been reported for 
the active K5 V star LO Peg (Jeffries et al. 1994).
 
$H_\beta$ emission in late type stars is another indicator of enhanced 
chromospheric activity. Figure 8 shows the spectra of FR Cnc and the comparison 
star HD 26794 in the $H_\beta$ region.  $H_\beta$ is seen clearly in emission,
while it is in absorption in the comparison star. The ratio of excess emission 
$\frac {E_{H_{\alpha}}}{E_{H_{\beta}}}$ with the correction given by Hall
\& Ramcey (1992) was also calculated for FR Cnc as 

$\frac {E_{H_{\alpha}}}{E_{H_{\beta}}} = \frac{EW(H_{\alpha})}{EW(H_{\beta})}*0.2444*2.512^{(B-R)}$\\

\noindent
This yields a value of 5.34 for the mean values of the equivalent widths of 
$H_{\alpha}$ and $H_{\beta}$. According to Hall \& Ramcey (1992), values of 
$\frac {E_{H_{\alpha}}}{E_{H_{\beta}}} > 3$ are indicative of the line emission 
likely arising from an 
extended region viewed off the limb, perhaps a stellar `prominence'.
 
 Figure 9  shows a comparison of spectrum of FR Cnc with that of HD 26794
in the region near the Ca II H and K lines. One can see that these lines have 
strong central emission components in the spectra of FR Cnc due to its 
active chromosphere. 
The spectrum of the Ca II H and K region taken on 15 November 2002 
($\phi =$0.72 and 0.76) was obtained close to the photometric epoch 'e'. 
The photometric light minima during this epoch are at phase of 0.44 and 0.75.
The equivalent widths of the Ca II H and K lines in the 15 November 2002 
spectra were found to be at a maximum during the photometric light minimum. 
Such a correlation is seen quite commonly  among chromospherically 
active stars and is typically interpreted as evidence that, as in the solar
case, the plage
regions responsible for the Ca II emission are adjacent to the spotted
regions responsible for the photometric modulation.

The equivalent widths of the $H_{\alpha}$ and Ca II H and K lines in FR 
Cnc are in the range of
1.2 to 1.8 \AA ~and 4.2 to 6.4 \AA, respectively. This is similar to the 
equivalent width found in most BY Dra type stars (e.g., see Bopp 1987).

\section{X-ray Spectra}
\label{xray.sec}

The X-ray spectra of HD 95559 and LO Peg are shown in 
the right panel and the left panel of  Figure 10, respectively. We created  
response matrices based on the available off axis calibration of the PSPC and 
using the appropriate ancillary response files. We used the {\it xspec}
(version 11.2) spectral analysis package to fit the data with spectral 
model for thermal equilibrium plasma known as the Mewe-Kaastra-Liedahl or MEKAL
model (Liedahl, Ostreheld \& Goldstein 1995; Mewe, Kaastra \& Liedahl 1995).
The background subtracted X-ray spectra were fitted with 1T and 2T plasma models
assuming the solar photospheric abundances  given in Anders \& Grevesse (1989)
and allowing the abundance of every element other than H to vary by a
common factor relative to their solar (photospheric) values. In each of the
above models we assumed that interstellar absorption follows the absorption
cross-sections given by Morrison \& McCammon (1983) and we allowed the 
total intervening hydrogen column density $N_H$ to vary freely.
We have also fitted the X-ray spectra with 1T and 2T collisional
plasma model {\it apec} (Smith et al. 2001). The fitted parameters 
using these models were similar to those determined using the MEKAL model.
For the star LO Peg there were two datasets corresponding to two observations. 
The spectrum was 
accumulated from each dataset and fitted using the above mentioned model 
individually as well as jointly.  
The best fit parameters were determined using the $\chi^{2}$ minimization 
technique. 
These parameters were found to be similar in either case.
Table 4 summarizes the best fit values obtained for the various parameters
along with minimum $\chi_{\nu}^{2} = \chi^{2}/\nu$ ($\nu$= degrees of freedom), 
and the 90\% confidence error bars estimated from minimum $\chi^{2} + 2.71$. 

Single-temperature MEKAL models  with abundances fixed  to the solar values 
gave an unacceptably high value of $\chi_{\nu}^{2}$ for the both the stars --
HD 95559 and LO Peg. 
 However, single-temperature MEKAL models with 
abundances $ < 0.02$ times the solar abundances and with 
plasma temperatures in the range of  0.63-0.77 keV were found acceptable
for the star HD~95559, but not for LO Peg.
 Two-temperature plasma models with the abundances fixed to the solar value 
were not found to be  acceptable for both the stars.
Acceptable MEKAL 2T fits were achieved when the abundances were 
allowed to depart from the solar values for both the stars. 
The best fit two-temperature plasma models
with sub-solar abundances along with the significance of 
the residuals in terms of their ratio for both the stars HD 95559 and LO Peg
are shown in the left and the right panels of the Fig. 10.
For HD 95559 an acceptable fit was obtained for abundances that were
 only $ 0.25_{-0.2}^{+0.3}$ times the solar values and with plasma components 
at temperatures of  0.45 keV and  1.27 keV. 
In the case of LO Peg the best fit plasma temperatures were 
 $0.30_{-0.05}^{+0.04}$ keV  and  $ 1.0_{-0.2}^{+0.1}$ keV, with abundances
that  were only  $ < 0.15$ times the solar values for both the
plasma components.

\section{X-ray and radio properties: comparison with similar systems}

A sample of 35 BY Dra systems were studied spectroscopically with 
ROSAT by Dempsey et al. (1997) assuming that their coronae had solar abundances.
The average value of the low temperature ($kT_{1}$) and the high 
temperature ($kT_{2}$) components for  these  BY Dra systems
are $0.19\pm 0.03$ keV and $1.31\pm 0.04$ keV, respectively.
The value of the $kT_{1}$ derived using sub-solar abundances 
for both the stars HD 95559 
and LO Peg (see Table 4) is found to be more than the mean value of the
BY Dra systems. Please note that the value of $kT_{1}$ derived using the solar
abundances (see Table 4)  is found to be consistent to that of the average value 
for the other BY Dra systems in Dempsey et al. (1997).
However, 2T plasma model using the solar abundances failed to fit the data 
with high signal-to-noise ratio and gave an unacceptable high value of $\chi_{\nu}^{2}$
for both of the stars.
The values of $kT_{2}$ are, however, consistent with that of the other BY Dra systems.
The average volume emission measures $EM_{1}$ and $EM_{2}$
for the BY Dra systems in  Dempsey et al. (1997) 
are $1.4\pm0.4 \times 10^{52}$ and $7.6\pm2.9 \times 10^{52}$ cm$^{-3}$, 
respectively. 
 The volume emission measure  $EM_{1}$ for the star HD 95559 is found to be 
$\sim 7$ times more than the average value of BY Dra systems. However, the value 
of $EM_{1}$ for LO Peg is consistent with that of the BY Dra systems.
The  value of $EM_{2}$ of the star HD 95559 using sub-solar abundances is consistent
with those of the other BY Dra systems.  However, the value of $EM_{2}$ for LO Peg is
$\sim 2-3$ times less than the average value of the other BY Dra systems.

X-ray flux measurements exist for a sample of 248 chromospherically 
active binary systems (CABS, Strassmeier et al. 1993).
On the basis of Simbad\footnote{The SIMBAD database is operated at CDS, Strasbourg, France}
spectral classification and the catalogue of CABS
we have divided the sample into 101 dwarfs, 65 subgiants, and 82 giants and  computed the 
average X-ray luminosity.  For the sample of  101 dwarfs the 
average and the median X-ray luminosity ($\rm{log}~L_{x}$) are $29.6 \pm 0.7$ and 
$29.63$ erg s$^{-1}$, respectively. 
The inferred $\rm{log}~L_{x}$ for HD 95559, LO Peg and FR Cnc are 30.3, 29.7, 
and $29.4 - 30.1$, where $L_{x}$ is in units of 
erg s$^{-1}$, respectively, which are all close to the  
average value for active dwarf systems.
The average and median X-ray luminosity for the 65 subgiants and, the 82 giants  active 
systems  are found to be $30.8\pm0.5$ 
and 30.83; $30.7\pm 0.7$ and 30.8 erg s$^{-1}$,
respectively.  This indicates
that the evolved systems (subgiants and giants) are more X-ray luminous than the
dwarf systems, as expected since these stars have larger surface areas from 
which to radiate X-rays. Radio flux measurements exist for a sample of 226 CABS 
(82 dwarfs, 64 subgiants and 80 giants). The average and median $\rm{log}L_{rad}
$ for a sample of 82 dwarfs are $14.8 \pm 0.7$ and 14.9 erg s$^{-1}$ Hz$^{-1}$, 
respectively.
LO Peg has been detected as a radio source of $(3.6 \pm 0.5) \times 10^{-26}$
$\rm{erg} \rm~{cm}^{-2} \rm{s}^{-1} \rm{Hz}^{-1}$ (Condon et al. 1998). 
However, FR Cnc and HD 95559 do not have any radio observations. Using a 
distance of 25.1 pc, the radio luminosity ($\rm{log}L_{rad}$) of LO Peg is 
found to be $15.43 \rm~{erg}~\rm{s}^{-1} \rm{Hz}^{-1}$, which is again close to 
the typical value for BY Dra type.

\section{Spectral Type, Temperature and Other Physical Parameters}

The  value of the total Galactic reddening $E(B-V)$ in the direction of FR Cnc
is estimated from Schlegel et al. (1998) to be 0.03 mag. However, given that
the Hipparcos parallax of this star implies a distance of only 33 parsec,
it is likely that its light suffers little if any reddening, and for the
remaining discussion, we assume $E(B-V) = 0.00$ mag.
Using the Hipparcos parallax of $30.24 \pm 2.03$ mas and the value of 
$V = 10.24$ mag from our 
photometry, the absolute magnitude $M_V$ of FR Cnc is 7.64 mag. (The Hipparcos
V-band measurements  and the V magnitude of Weis (1993) lead to similar
values for $M_{V} = 7.7$ mag
and $7.5$ mag, respectively). This value is consistent with 
a luminosity class V for this star. The color $(B-V)$
of FR Cnc derived from our photometry is $1.11 \pm 0.02$, compared to a redder,
but much lower precision, value of $1.62 \pm 0.20$ given in the Hipparcos
Catalogue. The 1.11 value of $(B-V)$ is best 
matched with a spectral class of K5 V. While this is three subclasses 
earlier than the spectral class (K8) given in the Hipparcos Catalogue,
the latter appears to be derived from a fairly crude spectral classification by 
Vyssotsky (1956) and thus the difference is likely not significant. 

We have determined the spectral energy distribution (SED) of FR Cnc,
HD 95559 and LO Peg using broad band UBVRI (present photometry and literature)
and 2MASS JHK (Cutri et al. 2003) fluxes. We have
assumed negligible reddening for each stars.
The observed  SED of FR Cnc, HD 95559 and LO Peg
are shown along with the synthetic SEDs (Kurucz 1993).
The synthetic SEDs are predicted from the intrinsic properties
of the stars. 
The model SEDs are shown in Figure 11.

The values of $T_{eff}$ and log g which best match to the observed SED
 are $4250 \pm 250$ K and $4.5 \pm
0.5$, respectively, and are consistent with the previously
inferred K5 V spectral type for FR Cnc.
The SED of HD 95559 (solid squares) is well matched with the synthetic 
SED for K1V type star. This 
indicates that HD 95559 is a K1V type star as concluded by Fekel \& Henry (2000).
 The SED of the star LO Peg is represented by solid triangles. 
Observations of the star LO Peg  by Jeffries et al. (1994) 
showed it to be a single K5-K7 dwarf, while Bowyer et al. (1996)
had reported it to be a K8 type star. From the present photometry
the absolute magnitude ($M_{V}=6.8$) and
color (B-V = 1.04) of the star LO Peg are consistent with the spectral 
type of K3V. It is also clear from the Fig. 11 that the synthetic SED of K3V 
($T_{eff}$ = $4750 \pm 250$;log g = 4.5)
type star is well matched with the SED of the LO Peg. 

Using  the Hipparcos parallax, the inferred $T_{eff}$ and log\ g values, and 
various relations given in Landolt-Borstein
(Schaifers \& Voigt 1982) we have determined values of 
$M/M_{\odot}, R/R_{\odot}, L/L_{\odot}$ for FR Cnc, HD 95559 and LO Peg.
These well-determined parameters along with the $M_V, M_{bol}$ are given 
in Table 5. 

\section{2MASS JHK Photometry of FR Cnc, HD 95559 and LO Peg}
One of the most remarkable properties sometimes attributed to chromospherically
active stars is the evidence of infrared (IR) excesses, which is related to 
the properties of diffuse circumstellar matter (Scaltriti et al. 1993).
Assuming negligible reddening and the 2MASS JHK magnitudes (Cutri et al 2003), 
we have determined the intrinsic $(\rm{J} - \rm{K})_0$ and $(\rm{H} - \rm{K})_0$
for FR Cnc, HD 95559 and LO Peg. The intrinsic $(\rm{J} - \rm{K})_0$ and 
$(\rm{H} - \rm{K})_0$
colors are 0.77 mag and 0.16 mag for a K5V type star and 0.54 mag and 
0.11 mag for K1V type star (Koornneef 1983). 

The intrinsic (J-K)$_0$  and (H-K)$_0$ colors of FR Cnc 
are  $0.74\pm0.03$ mag and $0.14\pm0.03$ mag imply a color 
excess of $0.03\pm0.05$ mag and  $0.02\pm0.03$ mag in (J-K) and (H-K) color,
respectively. 
The  intrinsic $(\rm{J} - \rm{K})_{0} = 0.539 \pm 0.03 $  and $(\rm{H}-\rm{K})_{0} 
= 0.111 \pm 0.03$ colors of HD 95559  give the 
color excess $0.00\pm0.03$ in each (J-K) and (H-K) color.
Assuming LO Peg as a K3V spectral type star and $(\rm{H} - \rm{K})_0 =0.69 \pm 0.03$
and $(\rm{J} - \rm{K})_0 = 0.14 \pm 0.03$. The color excess in LO Peg
is $0.02\pm0.04$ mag and $0.00\pm0.03$ mag, respectively.

The values of the color excess for each stars are consistent with 
zero to within the uncertainties, 
indicating that all these three stars  have no significant color excess in the JHK bands, 
which is also supported from the matching of the model SEDs as 
shown in Fig. 11.

\section{Kinematics}
We have computed the galactic space velocity components (U, V, W) for the star
FR Cnc from the proper motion and parallax measured by the Hipparcos and  the
radial velocities measured  by Upgren, Sperauskas \& Boyle (2001) as listed in Table 5. 
The resulting values and their associated errors are given in Table 6.

FR Cnc, HD 95559 and LO Peg are inside the boundaries for the young disk population in 
the (U,V) and (W,V) diagrams (Montes et al. 2001). The (U,V,W) components of
FR Cnc are close to that of the IC 2391 supercluster (-20.6, -15.7, -9.1) which is estimated to
have an age of 35 to 55 Myrs, and for which several dozen possible
late-type members have been previously identified. The star LO Peg has
also been identified as a young star previously: as either a member of the
100 Myrs old Local Association (Montes et al. 2001) or of the 50 Myr old 
AB Dor moving group (Zuckerman, Song \& Bessel 2004). 
We have also calculated the UVW component for the star HD 95559 using the
radial velocity $3.81\pm0.11$ km s$^{-1}$ (Fekel \& Henry 2000) and the 
parallax and proper motion given in the Hipparcos Catalogue. The UVW components 
of this star (see Table 6) indicate that it is possible member 
of the 600 Myr Hyades supercluster, although as already noted its strong
Li I absorption lines are more consistent with a Pleiades-type age of
100 Myrs.

\section{Summary and Conclusions}
The shape and amplitude of the photometric light curves of FR Cnc, HD 95559
and LO Peg are observed to be changing from one epoch to another. 
The change in the amplitude is mainly due to a change in the minimum of 
the light curve, and this may be due to a change in the spot coverage.   
This indicates the BY Dra type of variability in these dwarf systems. 
In such late-type of dwarfs, convection and rapid rotation generate a
dynamo, resulting starspots activity. 
Two groups of spots are identified for FR Cnc and LO Peg.
The spots are found to migrate, and migration periods of 0.97 year and 
0.93 year are determined from the 4 years of data.
A  migration period of 1.12 years for one group of spots in LO Peg
is also determined.   A new group of spots formation in the star
HD 95559 was seen during our observations.
We determine the spectral types of FR Cnc and LO Peg to be  K5V and 
K3V type, respectively. 
 
Long term  optical photometry of the star FR Cnc firmly establishes the
rotation period of the star to be $0.8267 \rm~{d}$.  
Variable $H_{\alpha}$, $H_{\beta}$  and Ca II H and K emissions in the 
spectra of the FR Cnc indicate the presence of a heated atmosphere
outside the photosphere. The chromospheric line emission seems to correlate 
with the photometric light curve, i.e. maximum at the light curve minimum, or
minimum at the light curve maximum. The excess emission line ratio of 
FR Cnc  implies that this emission likely arises from an extended region rather
than from plage or prominences.  
The X-ray spectral parameters of HD 95559 and LO Peg
are found to be consistent with that of other BY Dra type systems. 

The bulk of the optical, X-ray and kinematical data indicate that these 
three stars are all active, young stars of 100 Myrs or less, whose high
activity levels are primarily due to their youth. HD 95559 is particularly
interesting in that it is both young and a synchronized close binary system,
whose current rapid rotation is expected to persist for
many Gyrs, and it is remarkably similar to another young synchronized binary
system V824 Ara $=$ HD 155555 (see, e.g., Strassmeier \& Rice 2000).
We believe that this study has reconfirmed that selecting stars by their high
X-ray to optical flux ratios is an efficient way to identify young 
active stars in the solar neighborhood.

\acknowledgements
We are grateful to the referee Dr. Graham M. Harper for valuable comments and suggestions.
We are thankful to the time allocation committee  for giving time
at 234-cm VBT.
We are also thankful to Brijesh Kumar for useful discussions on the optical spectroscopic
aspects of the work.
This research has made use of data obtained from the High Energy Astrophysics Science
Archive Research Center (HEASARC), provided by NASA's Godard Space Flight center. 
Starlink is funded by PPARC and based at the Rutherford Appleton Laboratory, which is part 
of Council for the Central Laboratory of the Research Councils, UK.
This publication makes use of data products from the Two Micron All Sky Survey, which is a 
joint project of the University of Massachusetts and the Infrared Processing and Analysis 
Center/California Institute of Technology, funded by the National Aeronautics and Space 
Administration and the National Science Foundation.

\begin{deluxetable}{lccccc}
\tablecolumns{6}
\tablewidth{0pt}
\tablecaption{Observation log of the stars FR Cnc, HD 95559 and  LO Peg.}
\startdata
\hline
Object   & Nights  & Duration of  & Instrument   &Telescope &~~ \\
         &         &  observations   &              &          &   \\
\cutinhead{Optical Photometry}
FR Cnc  & 43~      &2001 Feb. - 2003 Jan.   &$2k\times2k$CCD &SNT &~~ \\
FR Cnc  & 10~      &2003 Dec. - 2004 Jan.   &$1k\times1k$CCD &SNT &~~ \\
FR Cnc  & 70*      &1990 Mar. - 1992 Nov.   &   \nodata &Hipparcos&Satellite\\
HD 95559 &  11~    &2001 Feb. - 2001 Apr~   &$2k\times2k$CCD &SNT &~~ \\
LO Peg   &  ~7~    &2001 Oct. - 2002 Oct.   &$2k\times2k$CCD &SNT &~~ \\
\cutinhead{Optical Spectroscopy} 
FR Cnc& 1 &2002 Nov. 15 &OMR {\normalsize spectrograph} &VBT & \\
      &   &                   &($1k\times1k$CCD) &   & \\
FR Cnc& 5 &2003 Jan. 20 - 24 & '' &VBT & \\
\cutinhead{X-ray observations from ROSAT}   
Source   & Observation   & Duration of   &exposure&Count rate   &offset\\
         &     ID        &  observations &(sec)   &(cts/sec)    & (arcmin)\\
HD 95559 & rp200987n00   &1993 June 2 - 9  &4874    &$0.62\pm0.02$&38.4   \\
LO Peg   & rp201753n00   &1993 Nov. 11 - 13 &5968    &$1.07\pm0.01$&0.31  \\
LO Peg   & rp201753a01   &1993 Dec. 7 - 8   &14012   &$1.02\pm0.01$&0.31   \\
\enddata
\tablenotetext{*}{Total number of measurements made by Hipparcos satellite}
\end{deluxetable}
\begin{deluxetable}{clcccc}
\tablecolumns{6}
\tablewidth{0pt}
\tablecaption{Photometry of FR Cnc, HD 95559 and LO Peg. }
\tablehead{
Mean epoch&Amplitude&$\Delta V_{max}$&$\Delta V_{min}$&\multicolumn{2}{c}{phase minima}\\
\cline{5-6}
240000.0+&~~~~$\Delta V$&&&I&II}
\startdata
\sidehead{\underline{FR Cnc}}
(a) 51984.75& ~~~~0.106  &-0.716&-0.610& 0.33 &0.75 \\
(b) 52245.50& ~~~~0.063  &-0.670&-0.607& 0.29 &0.66 \\
(c) 52267.00& ~~~~0.108  &-0.683&-0.575& 0.27 &~-~~ \\
(d) 52308.75& ~~~~0.100  &-0.697&-0.597& 0.23 &0.58 \\
(e) 52598.00& ~~~~0.153  &-0.725&-0.572& 0.44 &0.75 \\
(f) 52626.25& ~~~~0.158* &-0.738&-0.580& -    & -   \\
(g) 52643.75& ~~~~0.128* &-0.728&-0.600& 0.39 &0.62 \\
(h) 52992.00& ~~~~0.210  &-0.721&-0.511& 0.25 &0.54 \\
(i) 53022.25& ~~~~0.181* &-0.736&-0.555& 0.21 &0.53 \\
\sidehead{\underline{HD 95559}}
(a) 51945.40& ~~~~0.051  &-0.435&-0.384& 0.00 & -   \\
(b) 52003.80& ~~~~0.076  &-0.481&-0.405& 0.00 &0.53 \\
\sidehead{\underline{LO Peg}}
(a) 52190.80& ~~~~0.081  &-0.483&-0.402& 0.50 &-    \\ 
(b) 52548.00& ~~~~0.050  &-0.495&-0.445& 0.66 &-    \\ 
\enddata
\tablenotetext{*}{ Because certain phases were not covered fully during the 
observations, therefore minima and/or maxima at the epoch f, g and i could not 
be determined accurately}
\end{deluxetable}

\hspace{-2.5cm}
\begin{deluxetable}{lccccc}
\tablecolumns{6}
\tablewidth{0pt}
\tablecaption{$Ca II H\&K$, $H_{\alpha}$ and $H_\beta$ data of FR Cnc}
\tablehead{
JD&Phase&\multicolumn{4}{|c|}{EWs ($\AA$)}\\
\cline{3-6}
240000+&&$Ca II K$ &$H$ &$H_{\alpha}$ &$H_{\beta}$}
\startdata
52594.400&0.73&6.45&5.70&-    & -    \\
52594.431&0.77&5.39&4.58&-    & -    \\
52660.197&0.32& -   & -   &1.51& -    \\
52660.220&0.35& -   & -   &1.21& -    \\
52660.463&0.64& -   & -   & -   &0.40\\
52660.486&0.67& -   & -   & -   &0.33 \\
52661.204&0.54& -   & -   &1.498& -    \\
52661.251&0.59&5.19&4.59& -   & -    \\
52661.417&0.79& -   & -   &1.32& -    \\
52661.446&0.83& -   & -   & -   &0.36\\
52662.199&0.74& -   & -   &1.82& -    \\
52662.248&0.80&5.06&4.86& -   & -    \\
52662.379&0.96& -   & -   &1.39& -    \\
52662.452&0.05& -   & -   & -   &0.40\\
52663.168&0.91& -   & -   &1.55& -    \\
52663.217&0.97&5.67&4.22& -   & -    \\
52663.247&0.01& -   & -   &1.36& -    \\
52663.371&0.16& -   & -   &1.64& -    \\
52663.422&0.22& -   & -   & -   &0.31\\
52664.253&0.23& -   & -   &1.62& -    \\
52664.508&0.53& -   & -   & -   &0.31\\
\enddata
\end{deluxetable}

\clearpage
\begin{deluxetable}{lccccccccc}
\tablecolumns{10}
\tablewidth{-0pt}
\tabletypesize{\scriptsize}
\tablecaption{Results of X-ray spectral analysis. $\chi_{\nu}^{2} = \chi^{2}/\nu
$, where $\nu$ is degrees of freedom (DOF)}
\tablehead{
Object&Model&Abundances$^{a}$ &$N_H$           & $kT_1$    &$EM_1$          &$kT_2$ &$EM_2$          & $\chi_{\nu}^{2}$& DOF \\
      & &           &$10^{20}cm^{-2}$& (keV)     &$10^{52} cm{-3}$&(keV)  &$10^{52} cm{-3}$&       &     
}
\startdata
HD 95559&MEKAL 1T &1.0(fixed)& 0.00 & 0.31 &5.7&-&-&28.3&10\\
&MEKAL 1T &$ <0.02$&$ 0.6_{-0.4}^{+0.4}$ &$0.69_{-0.06}^{+0.08}$&$33.7_{-7}^{+8}$& -&-&1.7&9\\
&MEKAL 2T& 1.0(fixed)&0.00&0.26&4.1&1.16&5.8&2.5&8\\
&MEKAL 2T &$ 0.25_{-0.2}^{+0.3}$&$<0.45$ &$0.45_{-0.18}^{+0.19}$&$10.9_{-2.1}^{+3.1}$&$1.27_{-0.5}^{+4.4}$&$12.3_{-4.9}^{+3.1}$&1.6&7\\
LO Peg &MEKAL 1T  & 1.0(fixed)&0.00 &0.31                  &1.6                &      -           &  -  &37.7 &23\\
       &MEKAL 1T  & 0.01      &0.5  &0.58                  &7.6               &              -   &  -  &2.0  &22\\
&MEKAL 2T  & 1.0(fixed)&$<0.1$ &$0.21^{+0.01}_{-0.02}$&$0.91^{+0.06}_{-0.06}$&$ 0.85_{-0.04}^{+0.04}$&$0.93_{-0.04}^{+0.04}$&1.8&21\\
 &MEKAL 2T   &$ < 0.15 $&$< 4.1$&$0.30_{-0.05}^{+0.04}$&$2.5_{-0.6}^{+1.0}$&$ 1.0_{-0.2}^{+0.1}$&$1.9_{-0.1}^{+0.1}$&1.3&20\\

\enddata
\tablenotetext{a}{Common value of abundances for all the elements with respect to the solar photosphere values;}
\tablenotetext{~}{Notes: 1. Errors are with 90 \% confidence based on $\chi_{min}^{2} +2.71$; 
No errors or upper limits are derived when $\chi^{2}_{\nu}$ is $> 2$ }

\tablenotetext{~}{~~~~~~~~~~2. Distance = 54.3 pc (HD 95559), 25.1 pc (LO Peg).}
\end{deluxetable}

\clearpage
\begin{deluxetable}{llll}
\tablecolumns{4}
\tablewidth{0pt}
\tablecaption{Physical parameters of stars FR Cnc, HD 95559 and LO Peg}
\tablehead{
Parameter            &~~ FR Cnc           & HD 95559$^{*}$  &~~LO Peg }
\startdata
Sp. Type             &~~K5V               & ~~K1V  &~~K3V\\
$V$                  &~~$10.24\pm0.01$    & ~~$8.93^{a}$&~~$8.82\pm0.01$ \\
$M_{V}$(mag)         &~~$7.60 \pm0.01$    & ~~$6.04$    &~~$6.80\pm0.01$ \\
$M_{bol}$(mag)       &~~$6.83 \pm0.01$    & ~~$5.67$    &~~$6.30\pm0.01$ \\
$T_{eff}$($K$)       &~~$4250 \pm250$     & ~~$5250\pm250$  &~~$4750\pm250$ \\
log\ g               &~~$4.5  \pm0.5$     & ~~$4.5\pm0.5$   &~~$4.5 \pm0.5$ \\
$L/L_{\sun}$         &~~$0.13 \pm0.02$    & ~~$0.54\pm0.03$ &~~$0.25\pm0.02$\\
$M/M_{\sun}$         &~~$0.60 \pm0.02$    & ~~$0.81\pm0.01$ &~~$0.66\pm0.02$\\
$R/R_{\sun}$         &~~$0.70 \pm0.08$    & ~~$0.77\pm0.10$ &~~$0.72\pm0.10$\\
$\mu_{\alpha}$(mas/y)&$-98.93 \pm1.99^{a}$&$-140.79\pm1.24^{a}  $&~~$132.06\pm1.01^{a}$ \\
$\mu_{\delta}$(mas/y)&$-97.39 \pm1.44^{a}$&$~~4.91\pm0.87^{a}$  &$-144.83\pm0.93^{a}$ \\
$\pi$(mas)           &~~$30.24\pm2.03^{a}$&~~$18.43\pm1.19^{a}$ &~~$39.91\pm1.18^{a}$\\
$R_{V}$ (km/s)       &~~$27.0 \pm 2.3^{b}$&~~$3.81\pm0.11^{c}$  &$-17.4\pm1.0$ \\
                     &~~$24.0 \pm 4.0^{b} $&~~~~~\nodata&~~~~~\nodata \\
\enddata
\tablenotetext{~}{$^{*}$Parameters are determined for individual component of 
binary system, $^{a}$Hipparcos, $^{b}$Upgren, Sperauskas \& Boyle 2002, $^{c}$ Fekel \& 
Henry 2000}
\end{deluxetable}

\clearpage
\clearpage
\begin{deluxetable}{ccccc}
\tablecolumns{5}
\tablewidth{0pt}
\tablecaption{Galactic space-velocity components. All the units are in km s$^{-1}$}
\tablehead{
$R_{V}\pm \sigma_{R_{V}}$& $U \pm \sigma_{U}$ & $ V \pm \sigma_{V}$ & $W \pm \sigma_{W}$ & $V_{total}\pm \sigma_{V_{total}}$}
\startdata
\sidehead{\underline{FR Cnc}}
$27.0 \pm 2.3$ &$-25.2  \pm 1.7$ &$-23.4  \pm 0.9$  &$-4.3 \pm 1.4$&$34.7\pm2.0$\\
$24.0 \pm 4.0$ &$-23.0  \pm 3.1$ &$-22.2  \pm 1.7$  &$-5.7 \pm 2.2$&$32.5\pm3.7$\\
\sidehead{\underline{HD 95559}}
$3.81\pm0.11$&$-32.9\pm0.4$&$-10.9\pm0.2$ &$-11.2\pm$0.1&$36.4\pm05$\\
\sidehead{\underline{LO Peg}}
$-17.4\pm1.0$  &$-5.2\pm0.29$&$-23\pm0.95$&$-23.86\pm0.95$&29.0\\
\enddata
\end{deluxetable}
\clearpage
\begin{figure*}
\hbox{
\includegraphics[height=4.5cm,width=7cm]{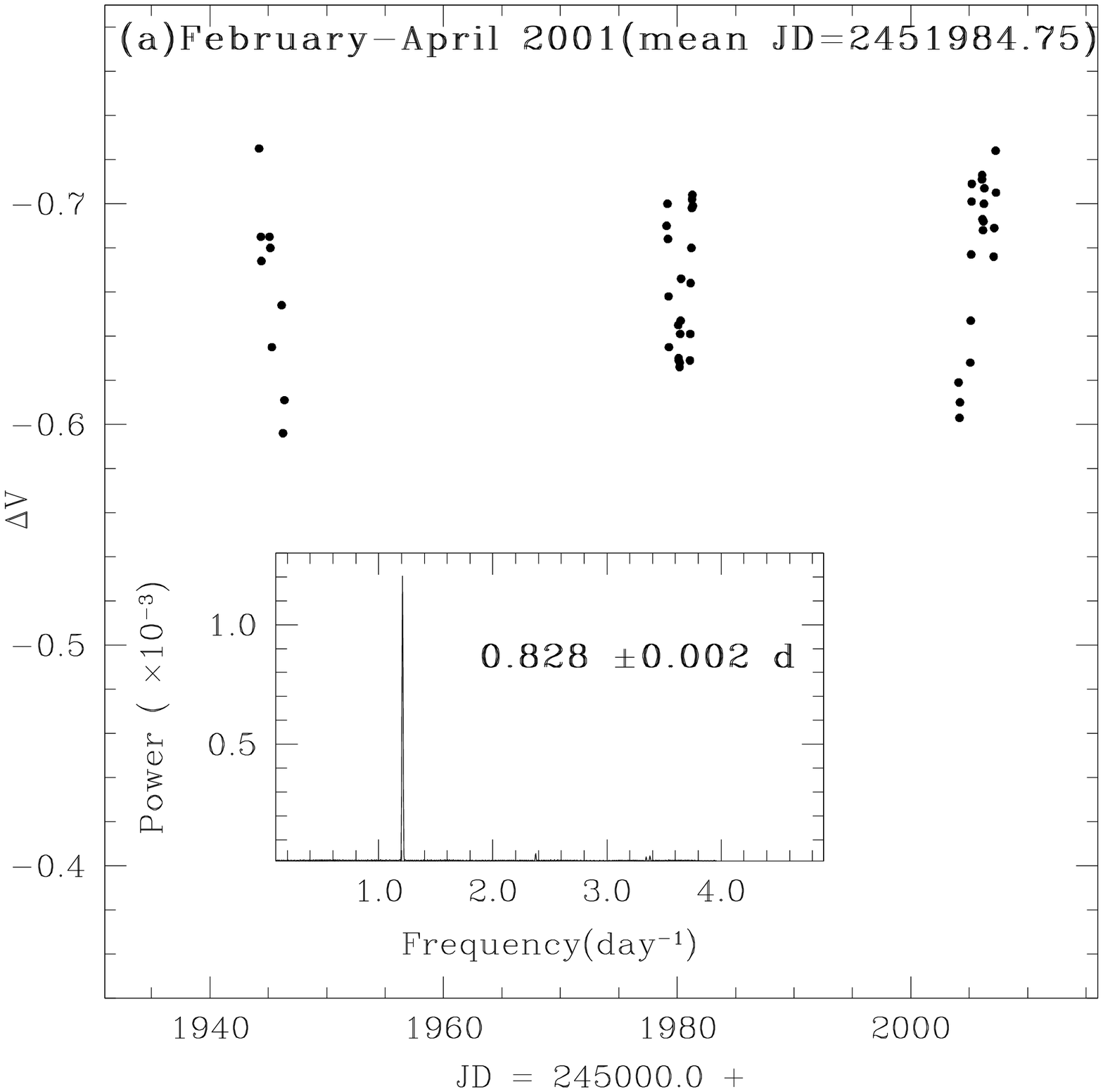}
\hspace{1.5cm}
\includegraphics[height=4.5cm,width=7cm]{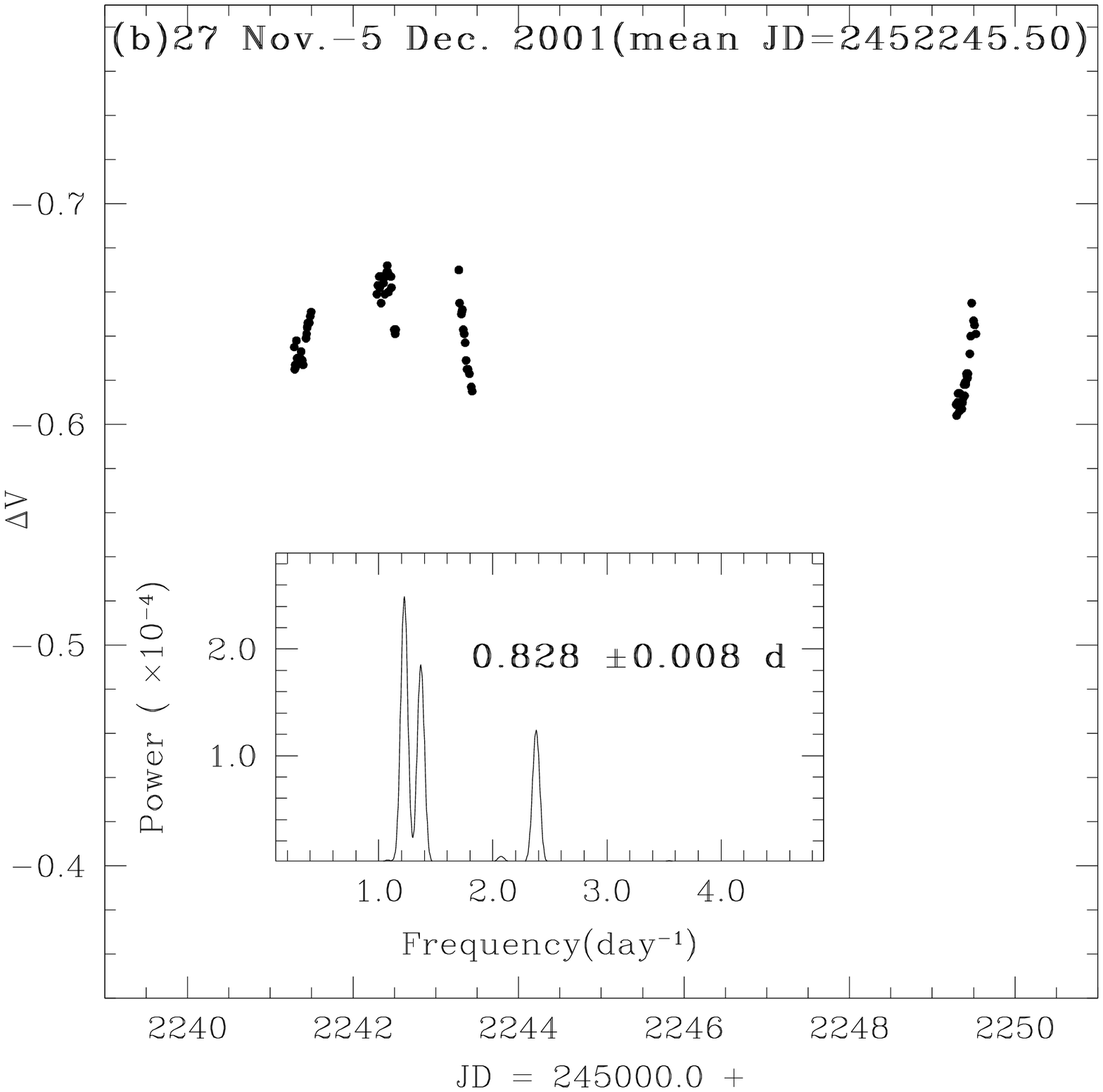}
}
\hbox{
\includegraphics[height=4.5cm,width=7cm]{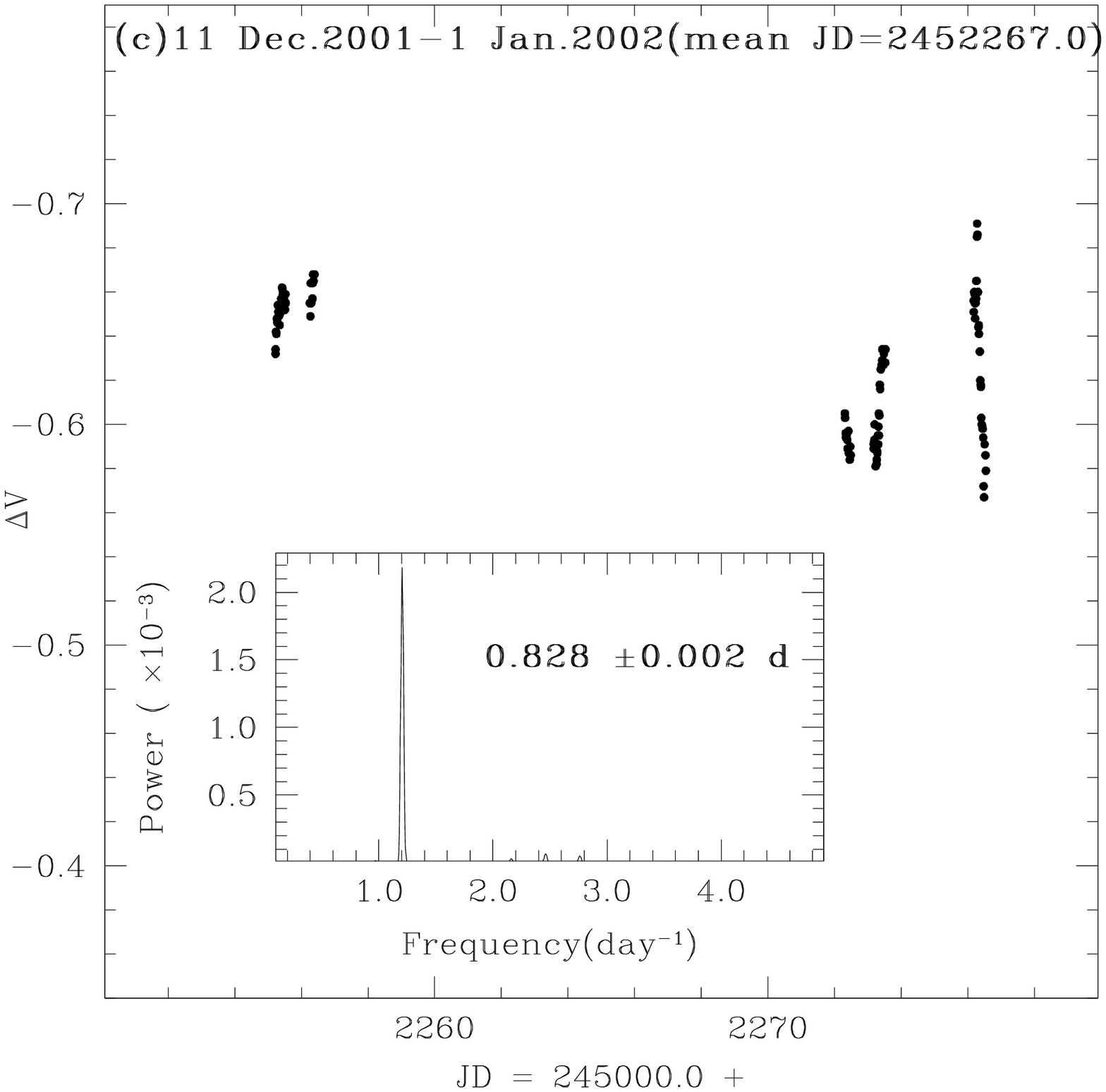}
\hspace{1.5cm}
\includegraphics[height=4.5cm,width=7cm]{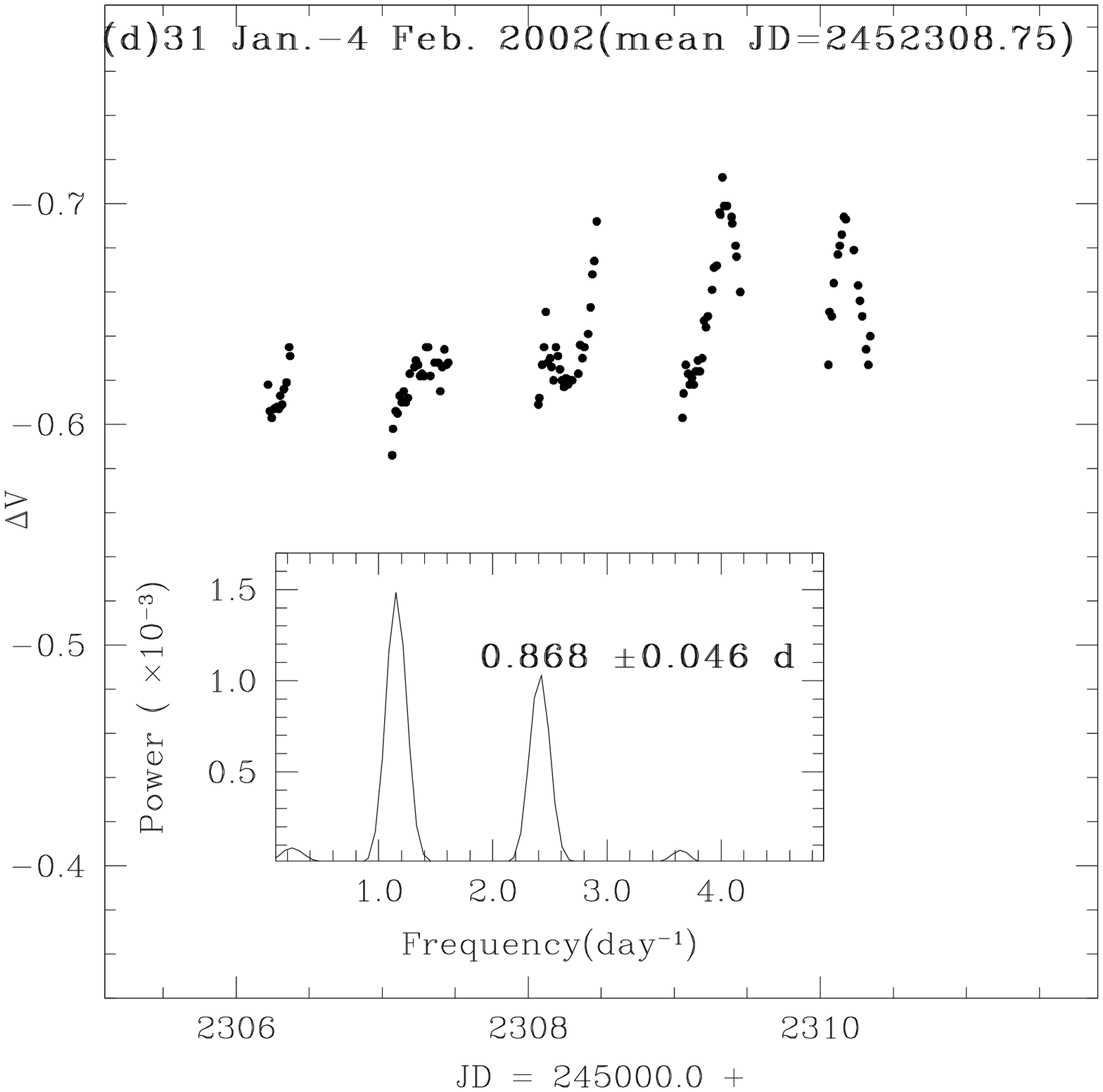}
}
\hbox{
\includegraphics[height=4.5cm,width=7cm]{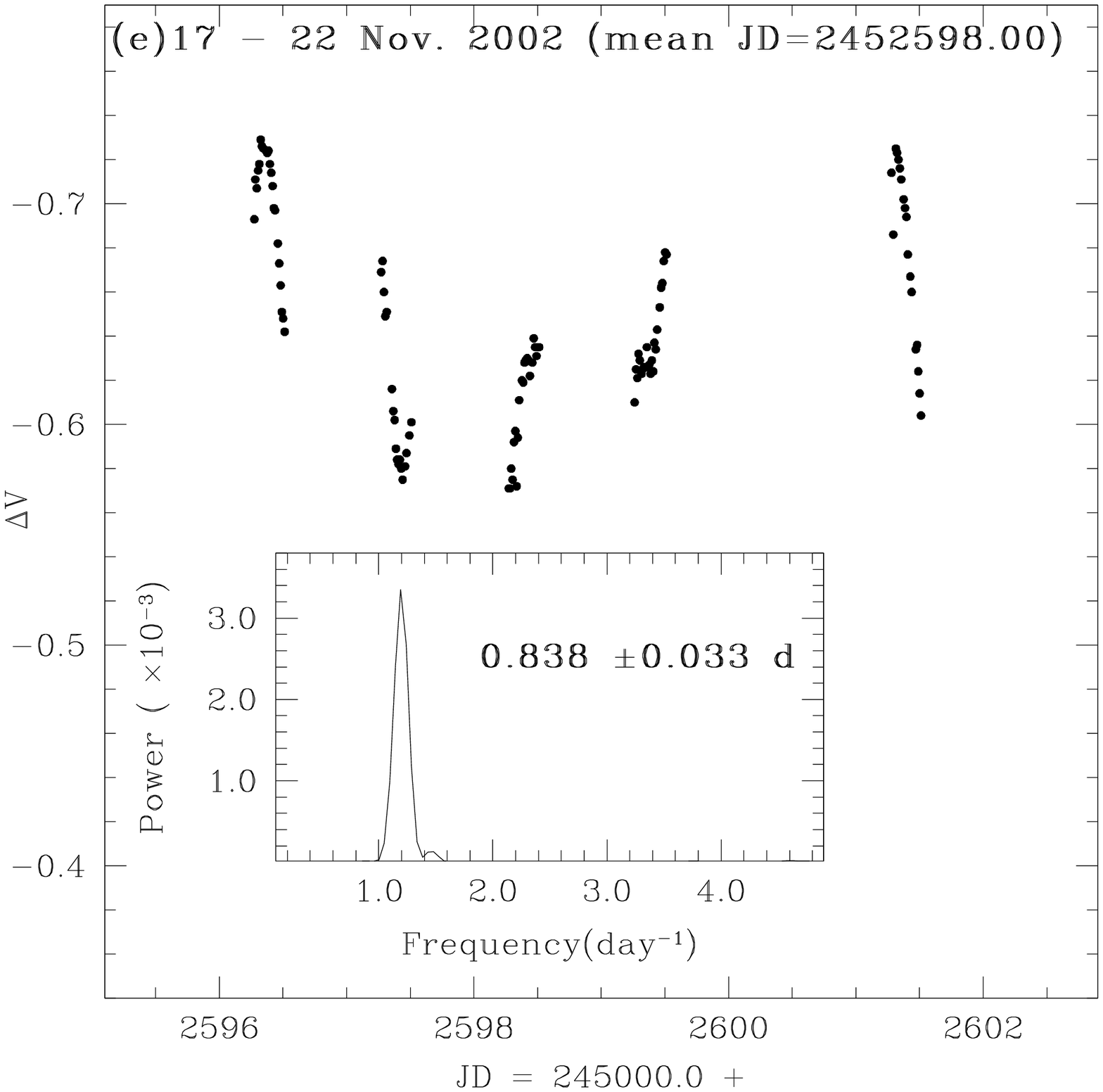}
\hspace{1.5cm}
\includegraphics[height=4.5cm,width=7cm]{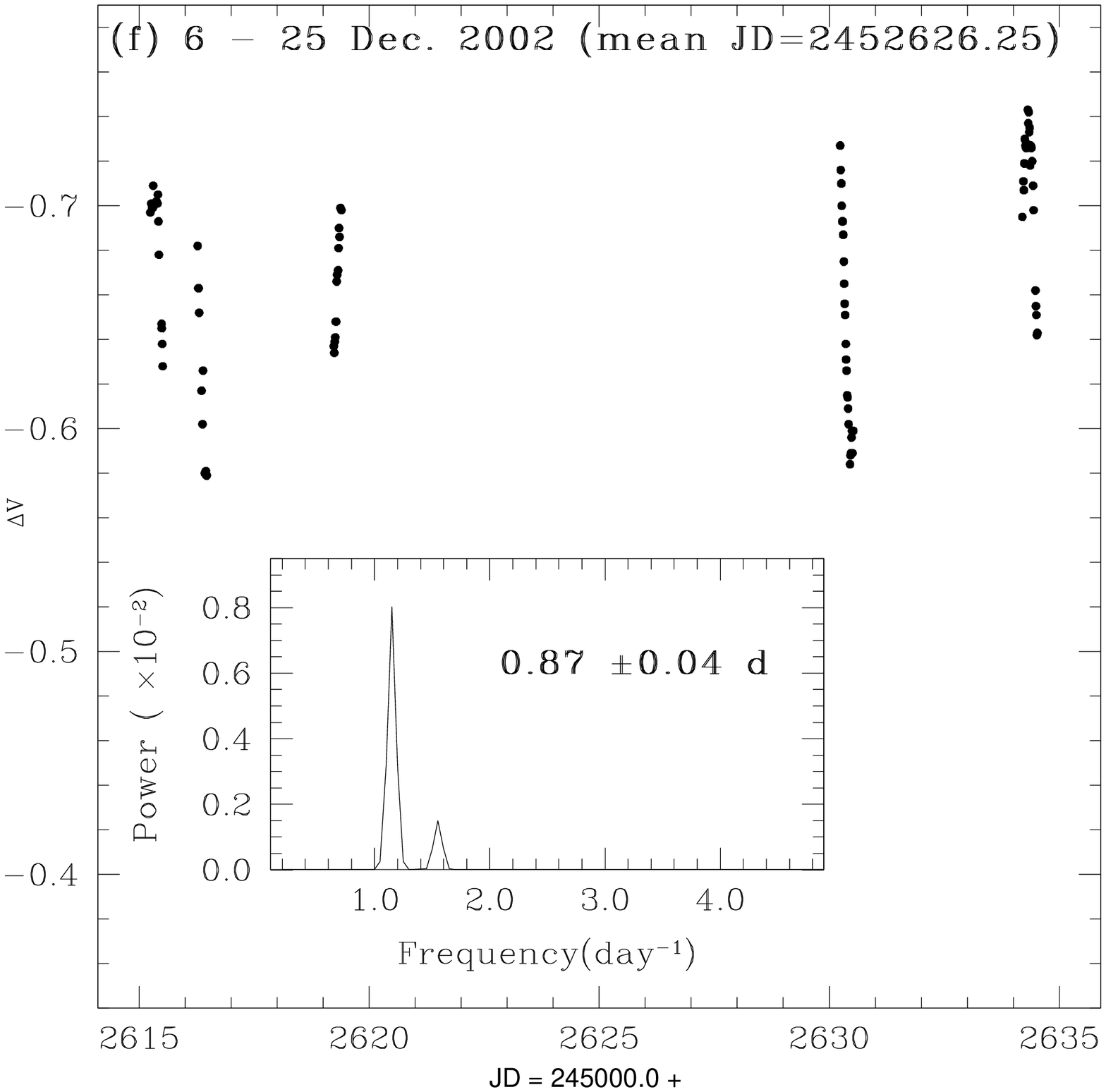}
}
\hbox{
\includegraphics[height=4.5cm,width=7cm]{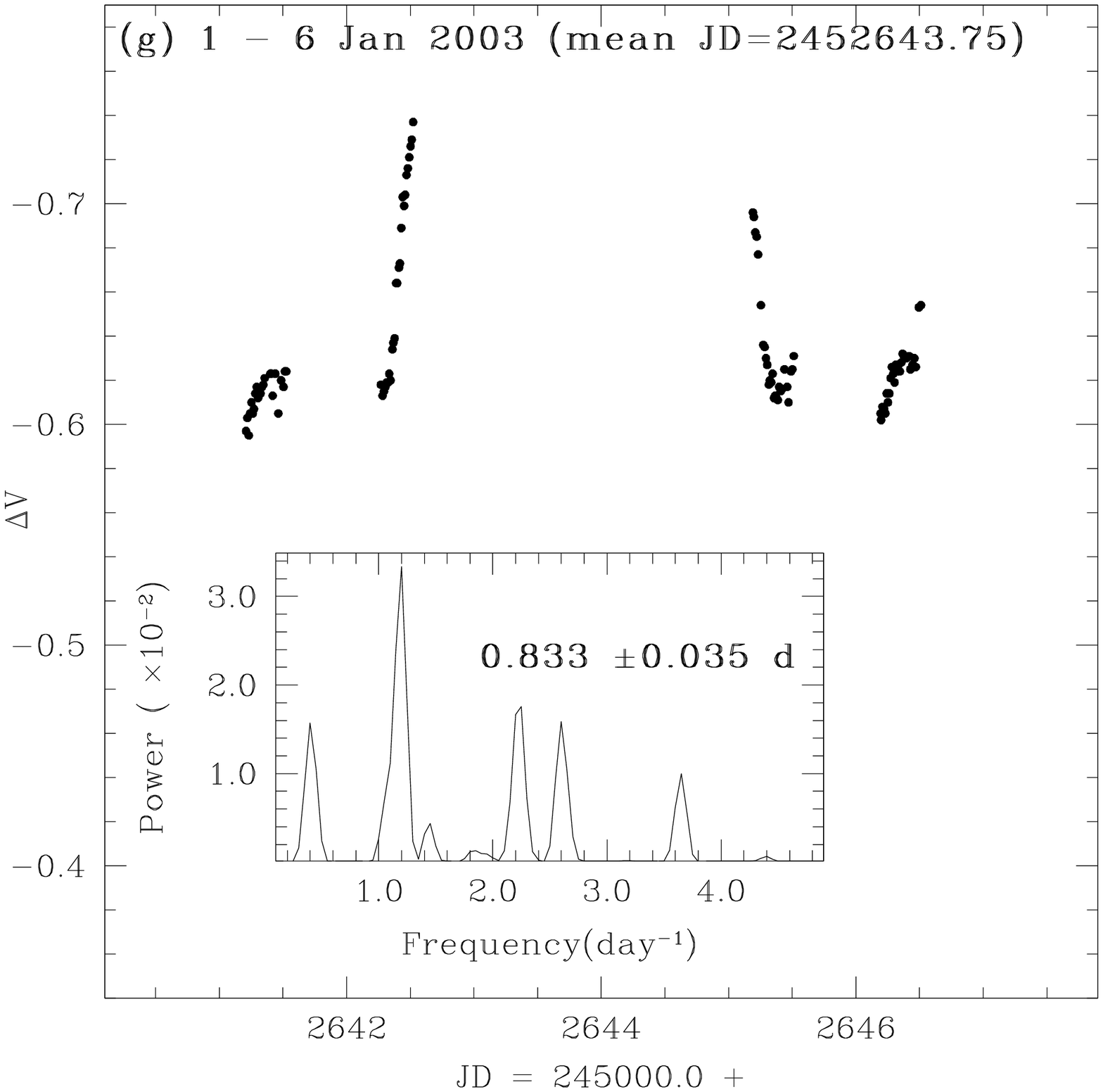}
\hspace{1.5cm}
\includegraphics[height=4.5cm,width=7cm]{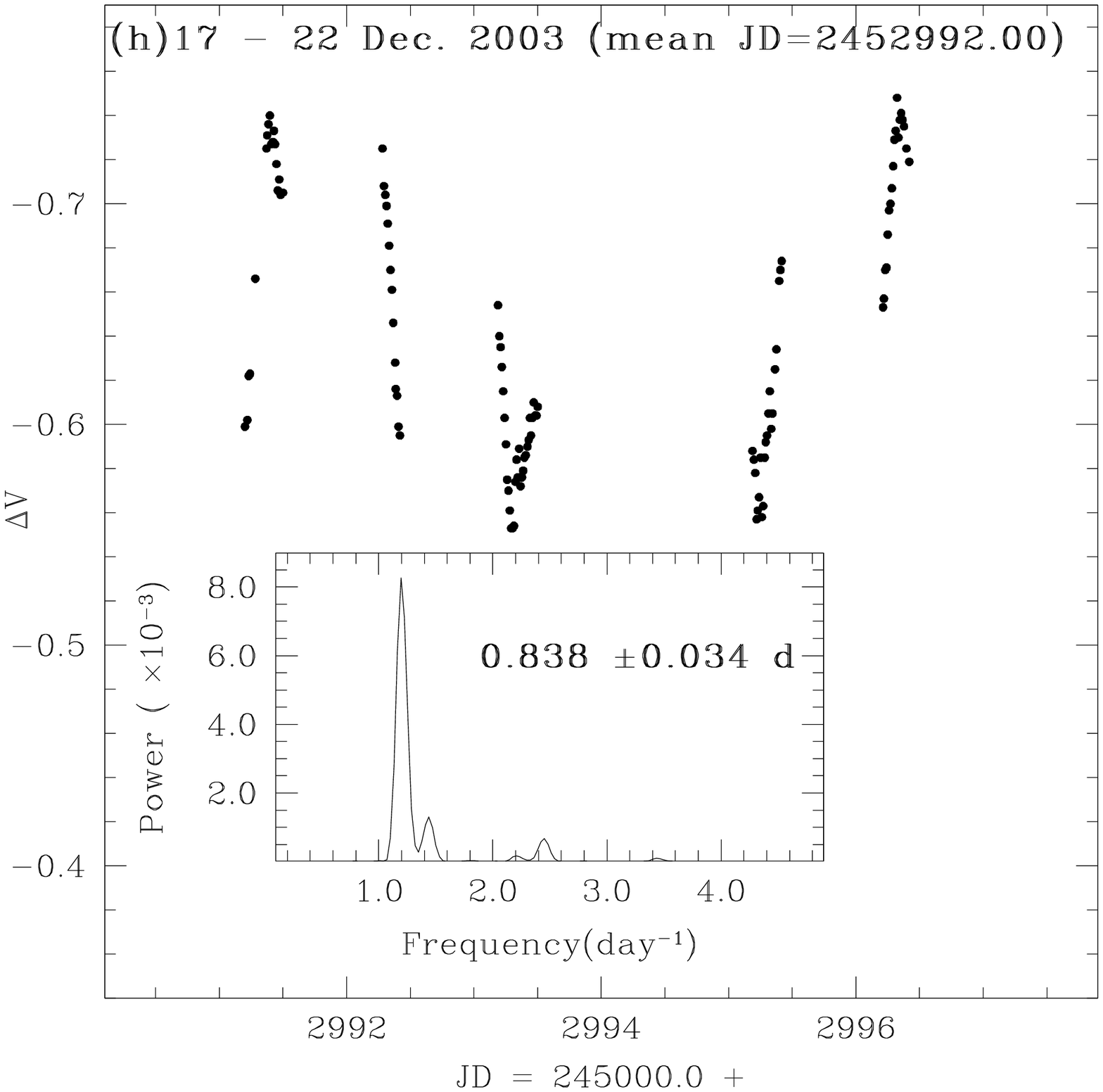}
}
\hbox{
\includegraphics[height=4.5cm,width=7cm]{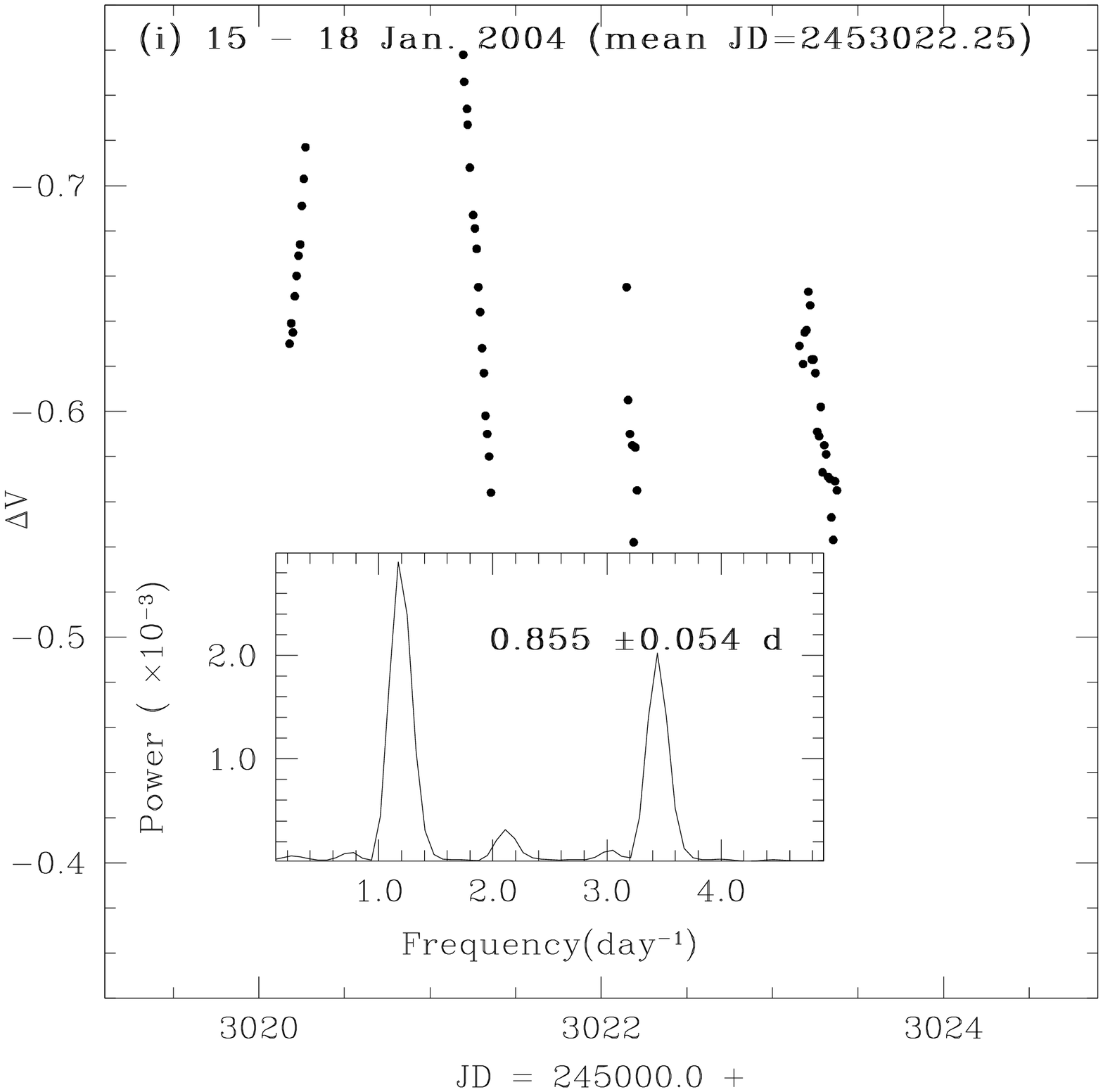}
\hspace{1.5cm}
\includegraphics[height=4.5cm,width=7cm]{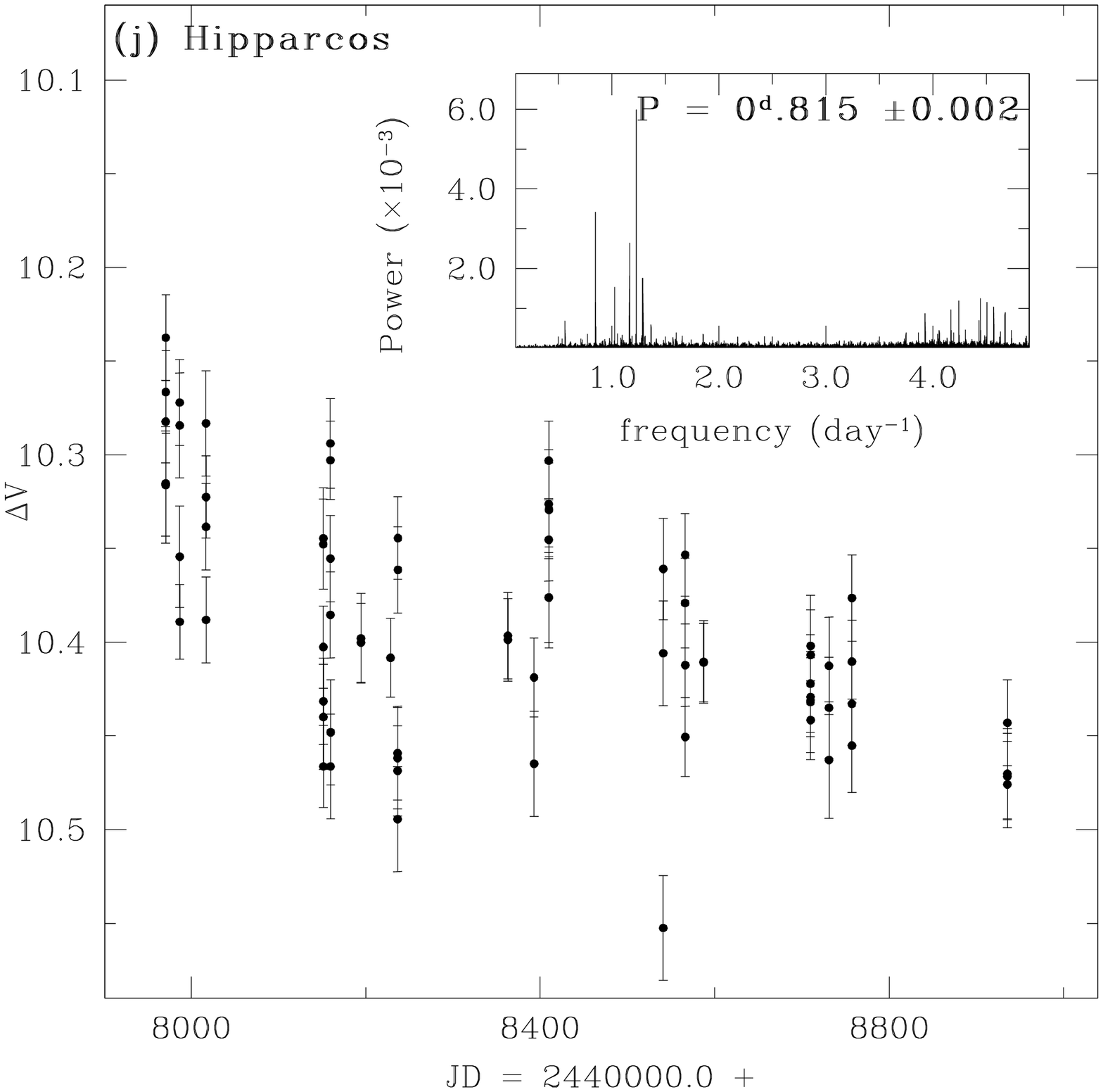}
}

\end{figure*}
\clearpage

\figcaption{Light curves and corresponding CLEANed power density spectra (insets) 
of FR Cnc at different epochs. The epoch is mentioned at the top of 
each panel, and the period is written at the top of  inset. 
The bottom right panel (j) shows the Hipparcos light curve along with its power 
spectrum}
\clearpage
\begin{figure}[h]
\hspace*{-1.0cm}
\hbox{
\includegraphics[height=10.0cm,width=16.0cm]{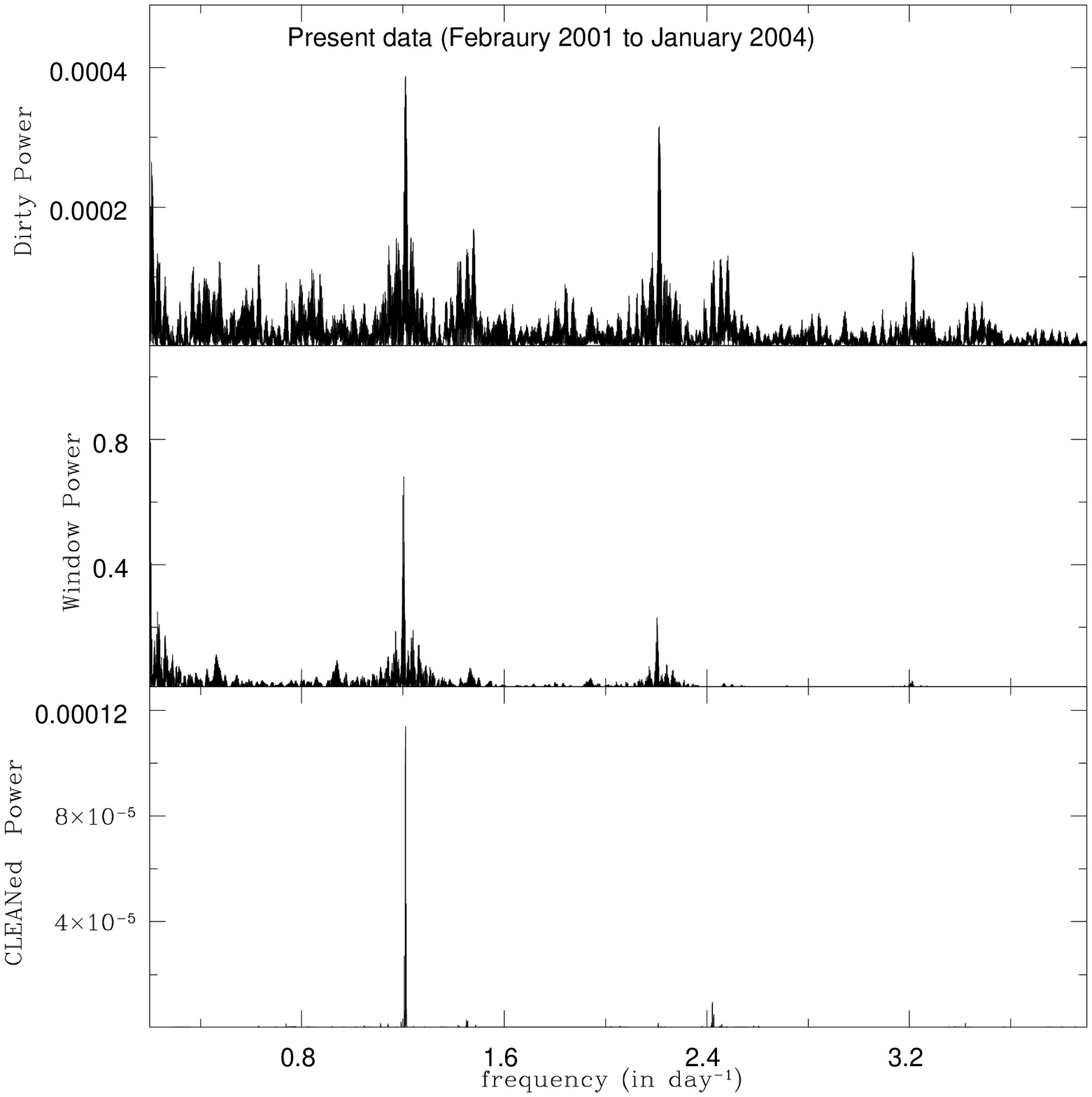}
}
\caption{Power spectra of FR Cnc from the entire photometric data 
taken during 2001-2004. {\it Top:} Dirty power density spectrum;
{\it Middle: } corresponding window power; {\it Bottom: } 
CLEANed power density spectrum}. 
\end{figure}
\clearpage
\begin{figure*}
\hbox{
\includegraphics[height=7.0cm,width=5cm]{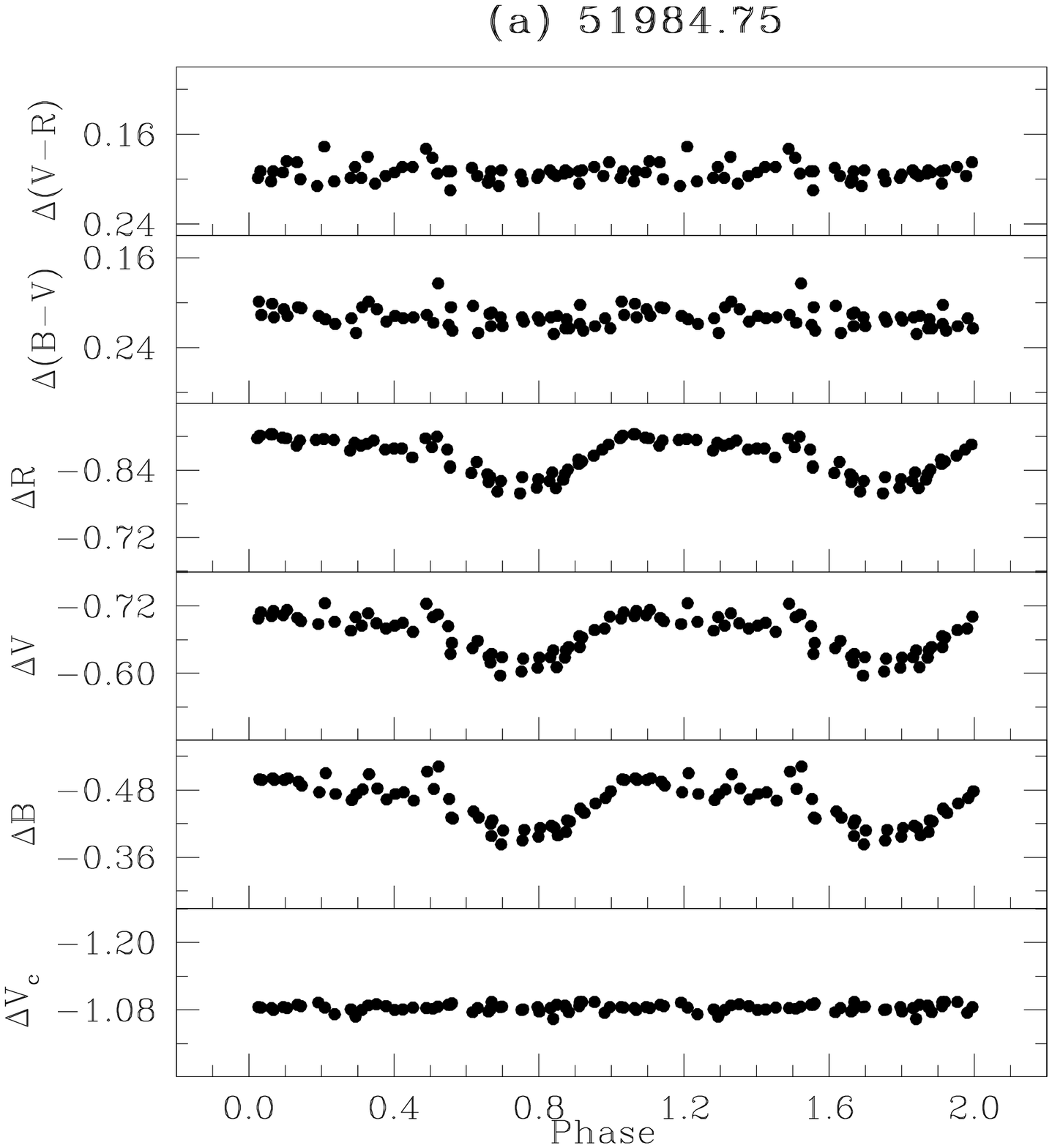}
\hspace{0.6cm}
\includegraphics[height=7.0cm,width=5cm]{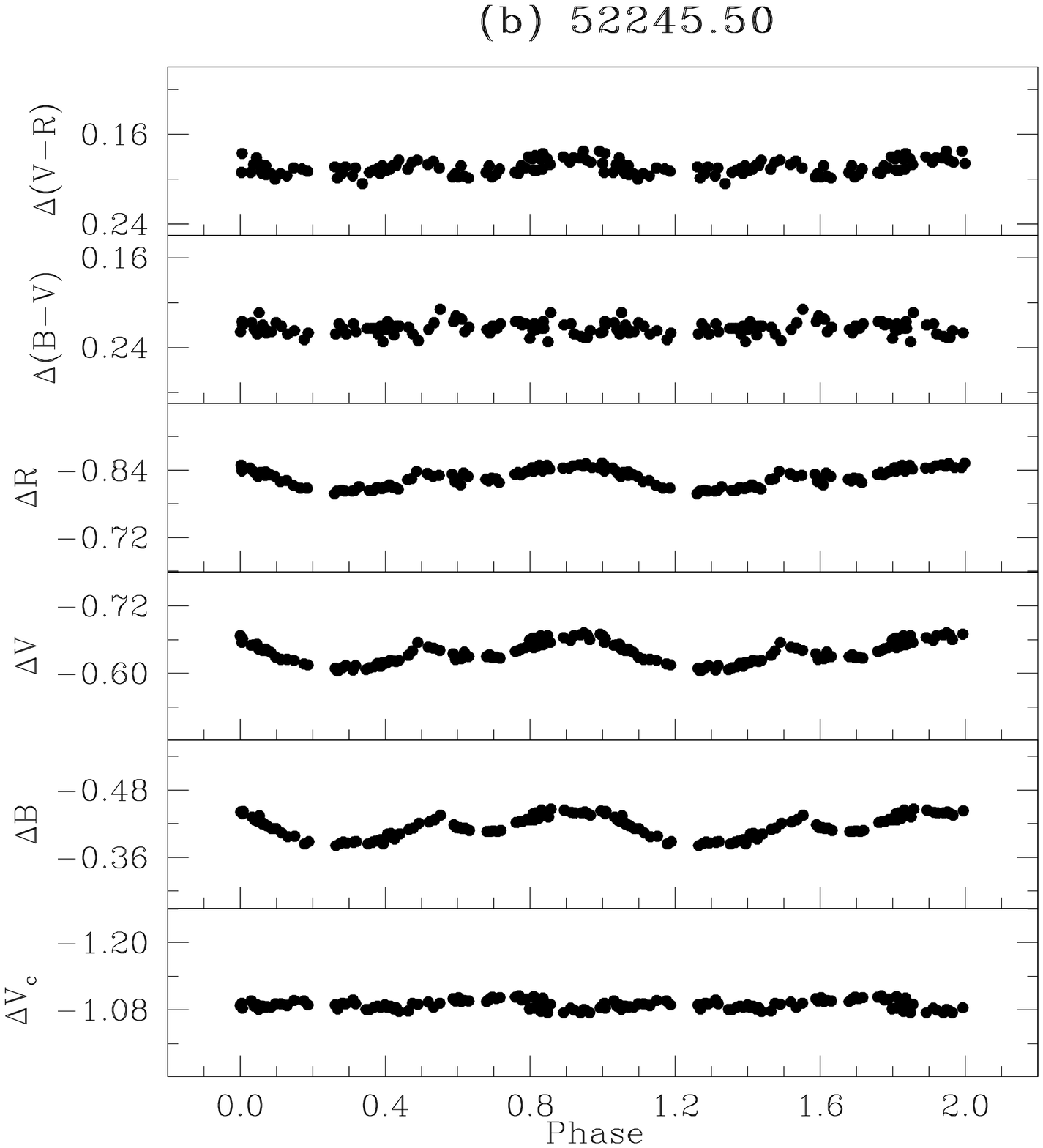}
\hspace{0.6cm}
\includegraphics[height=7.0cm,width=5cm]{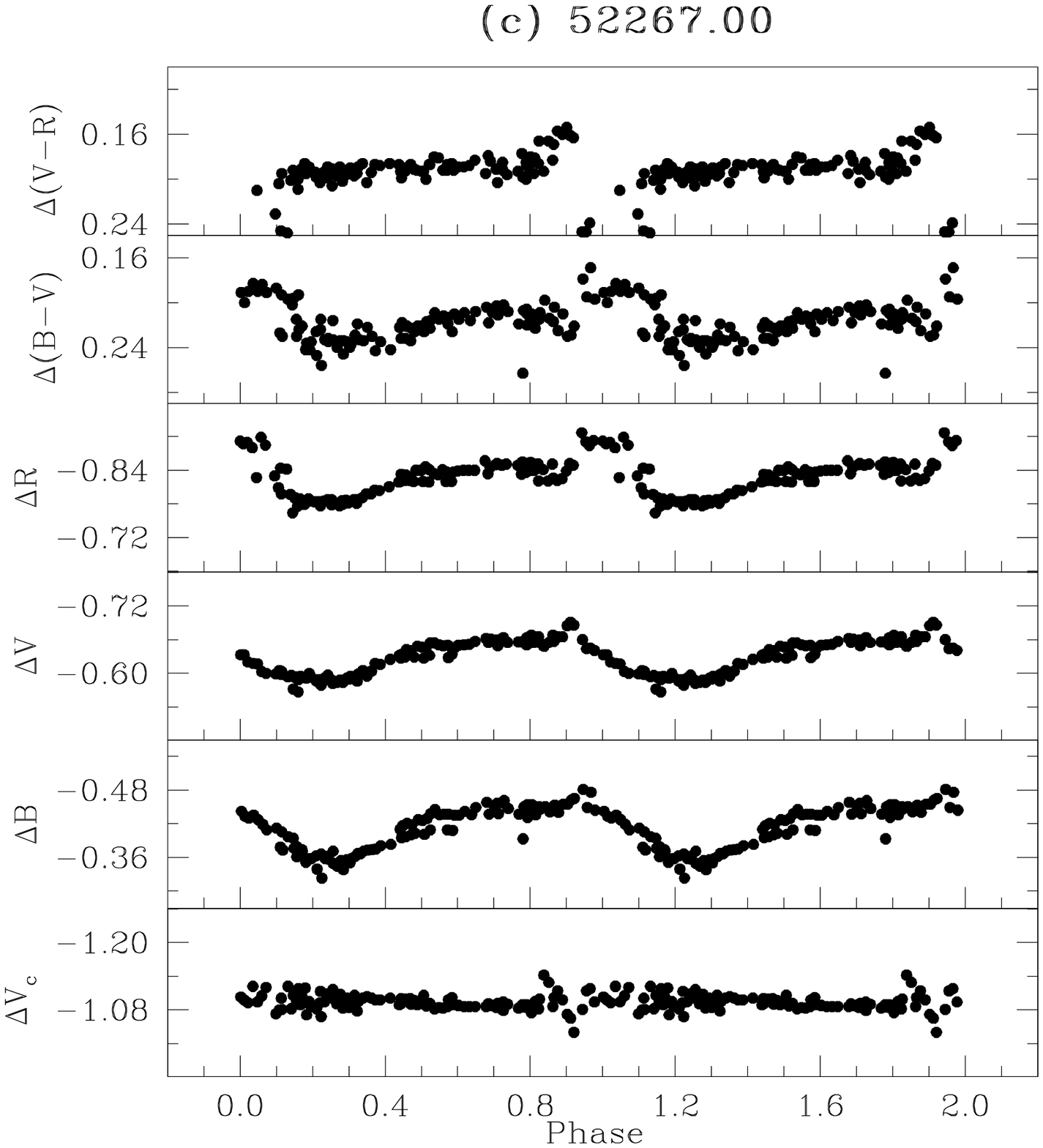}
}

\vspace{0.4cm}

\hbox{
\includegraphics[height=7.0cm,width=5cm]{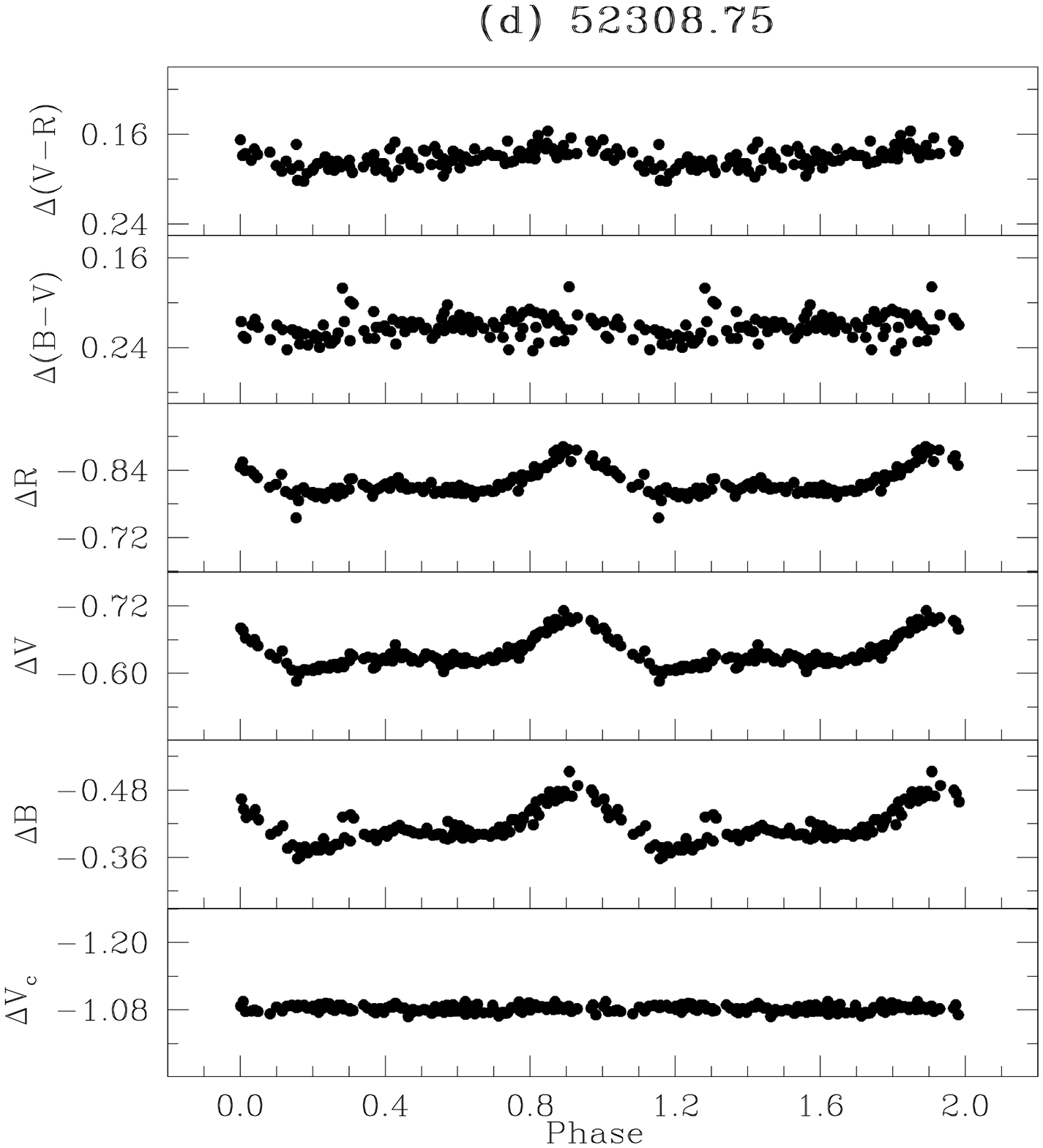}
\hspace{0.6cm}
\includegraphics[height=7.0cm,width=5cm]{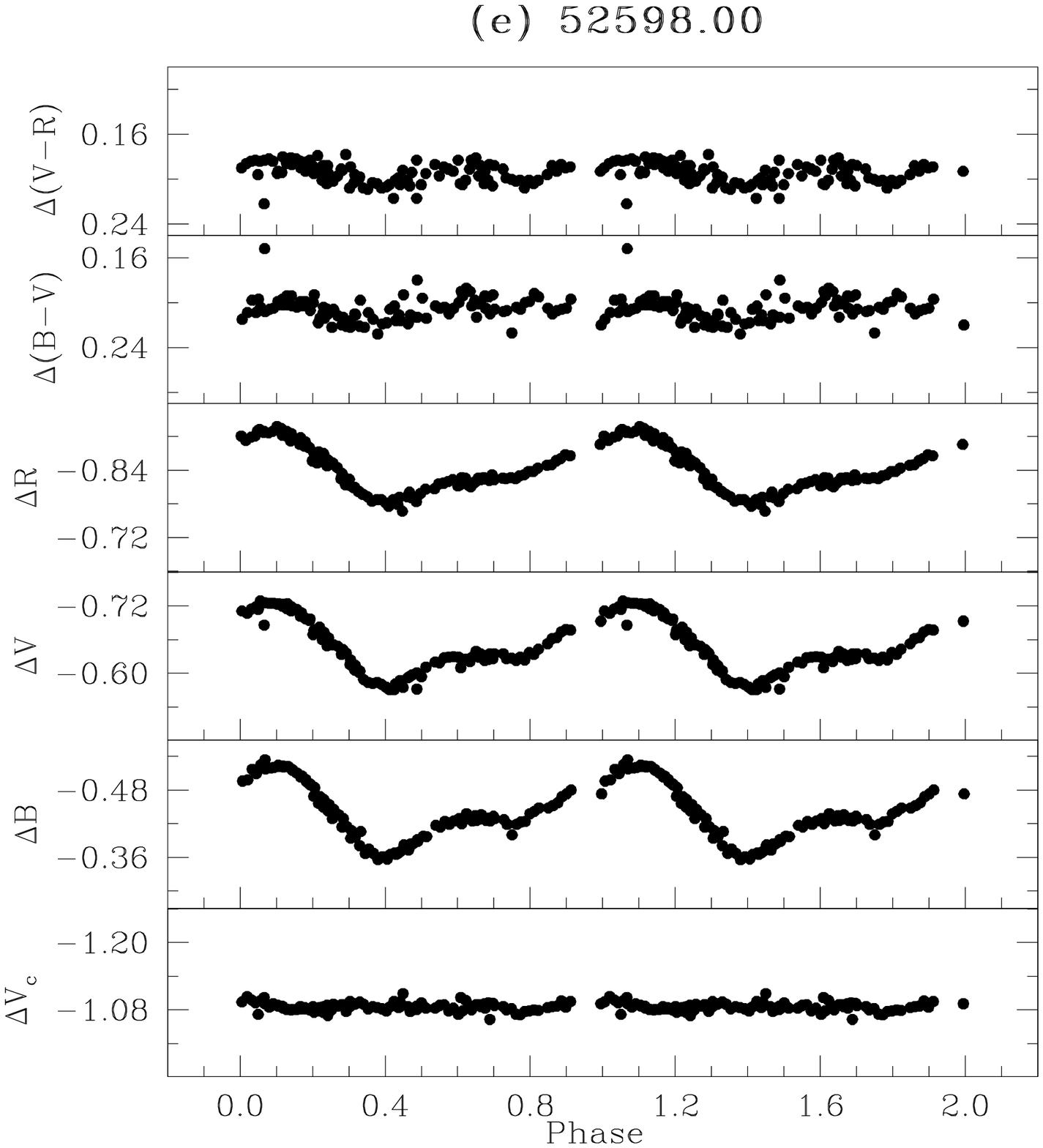}
\hspace{0.6cm}
\includegraphics[height=7.0cm,width=5cm]{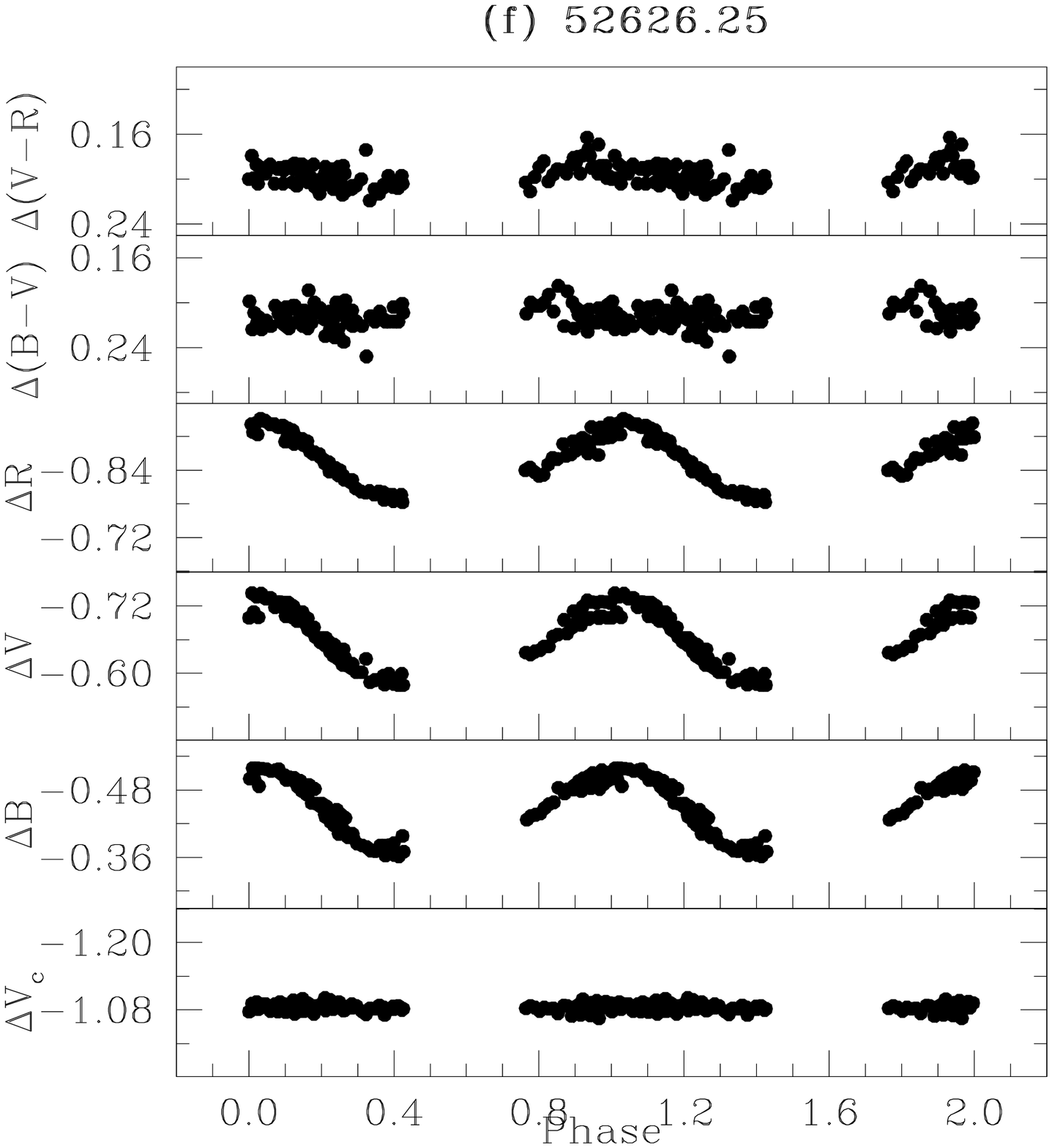}
}

\vspace{0.4cm}

\hbox{
\includegraphics[height=7.0cm,width=5cm]{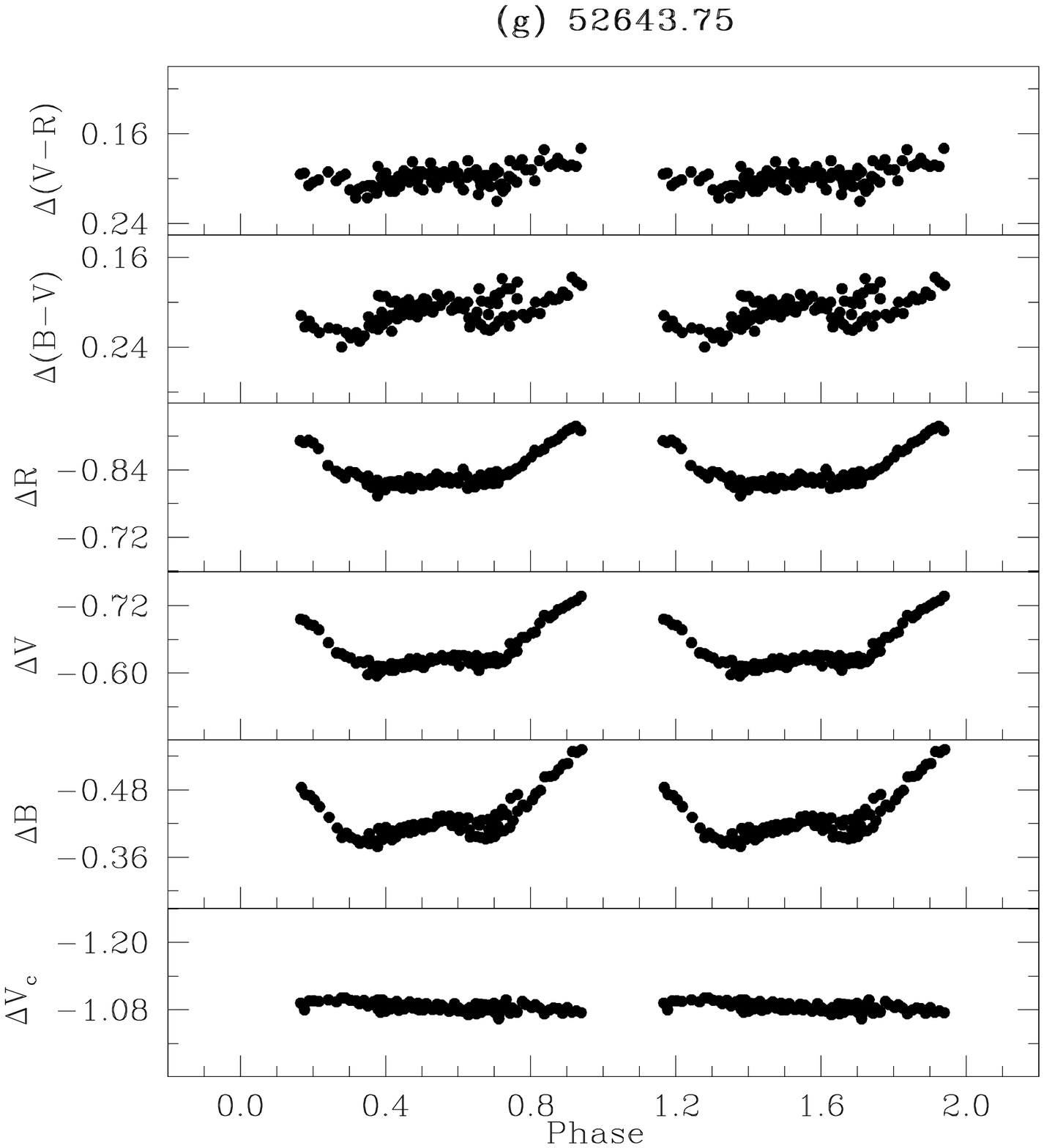}
\hspace{0.6cm}
\includegraphics[height=7.0cm,width=5cm]{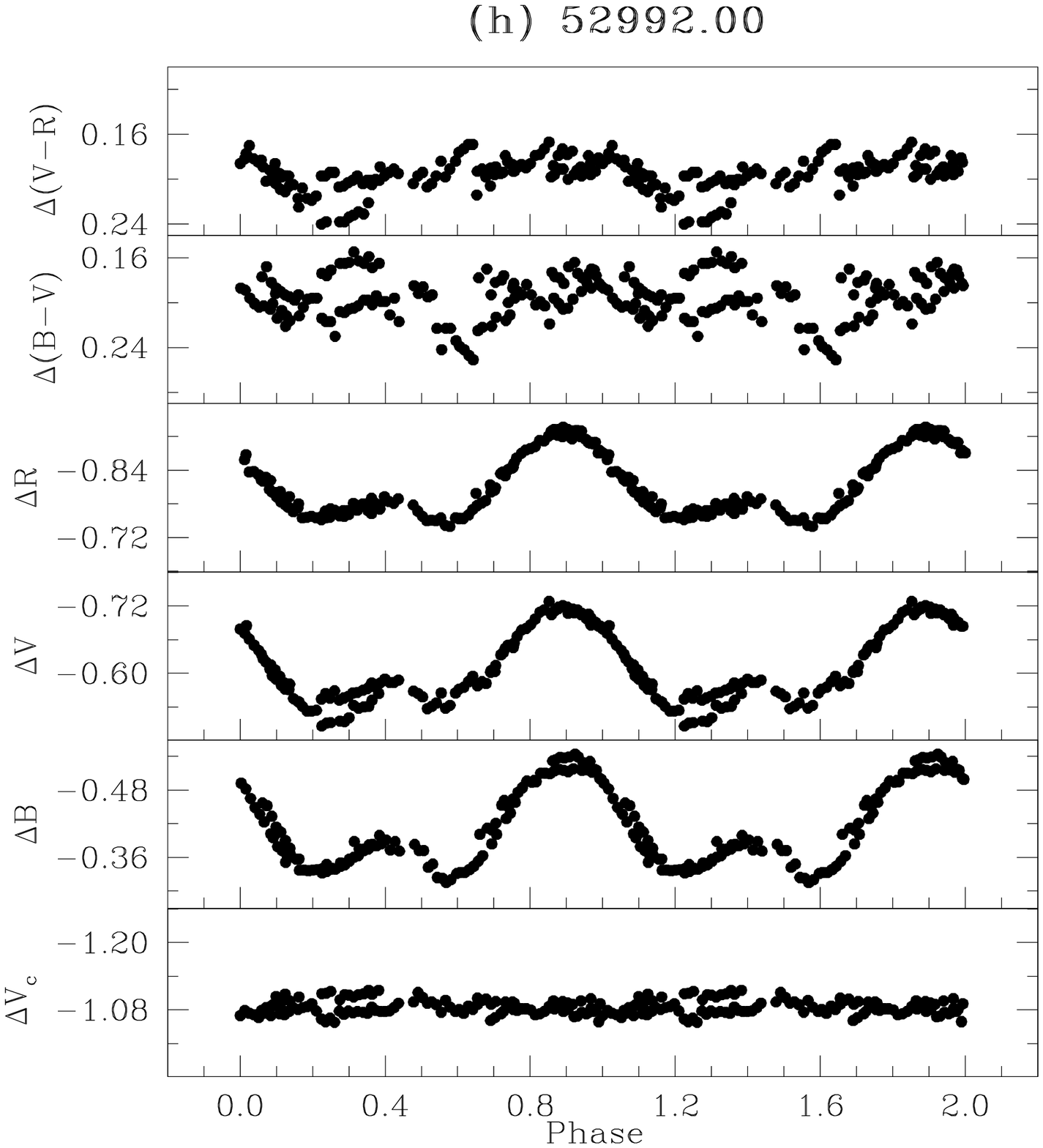}
\hspace{0.6cm}
\includegraphics[height=7.0cm,width=5cm]{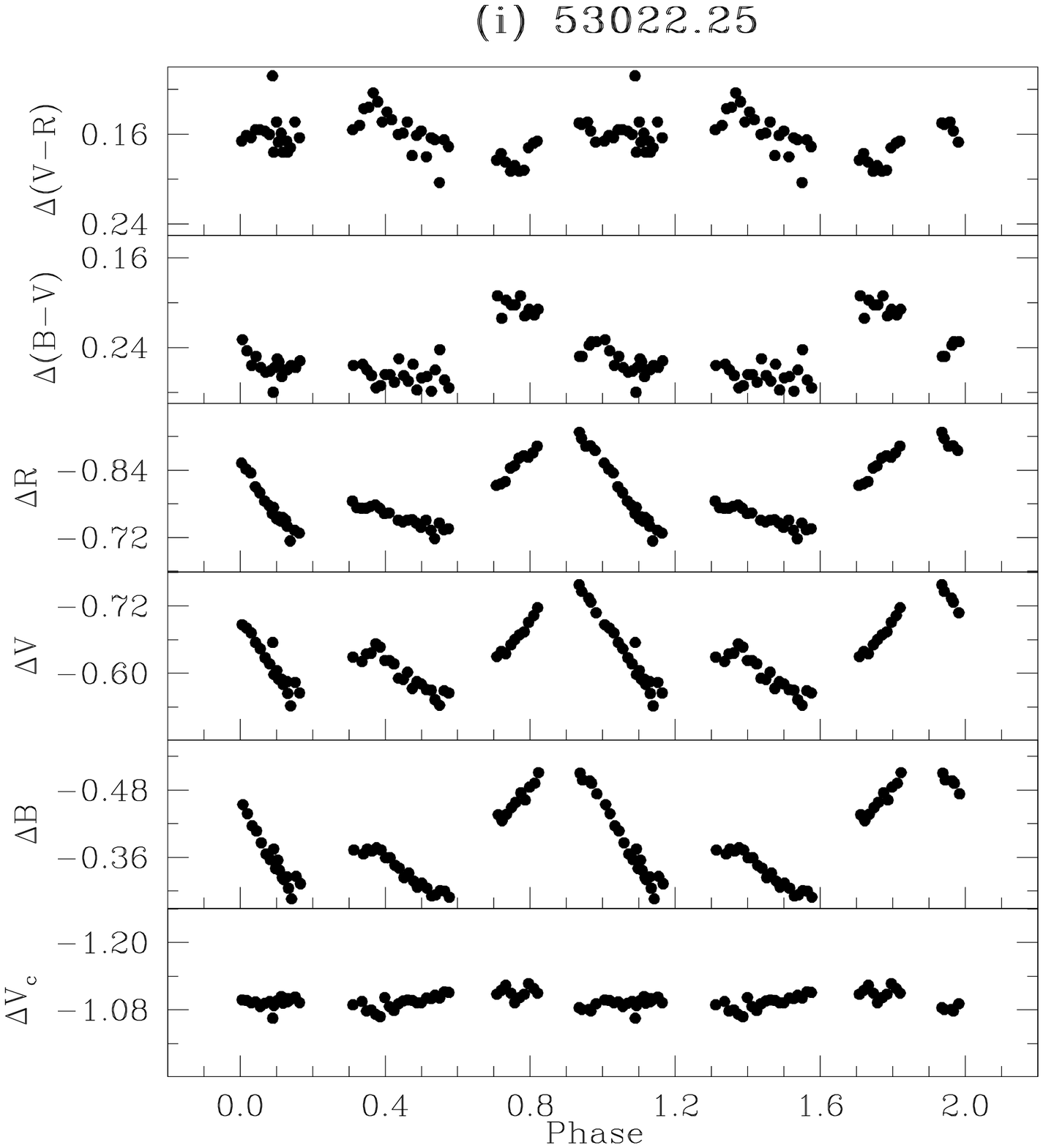}
}
\end{figure*}
\clearpage
\figcaption{$V_{c}, B, V, R $ light curves and (B-V), (V-R) color curves of 
FR Cnc folded using a period of $0.8267 \rm~{d}$, and shown for nine 
different epochs.  The epoch (JD=2400000.0+ is marked at the top of each panel.
The bottom panel in each Figure represents the plot of different measures
between the comparison (TYC 1392 2120 1 for Fig. a to g and BD +16\degr 1751 
for Fig h and i)
and the check (USNO-A2.0 1050-05766589) star.}
\clearpage
\begin{figure}
\hbox{
\includegraphics[height=8cm,width=8cm]{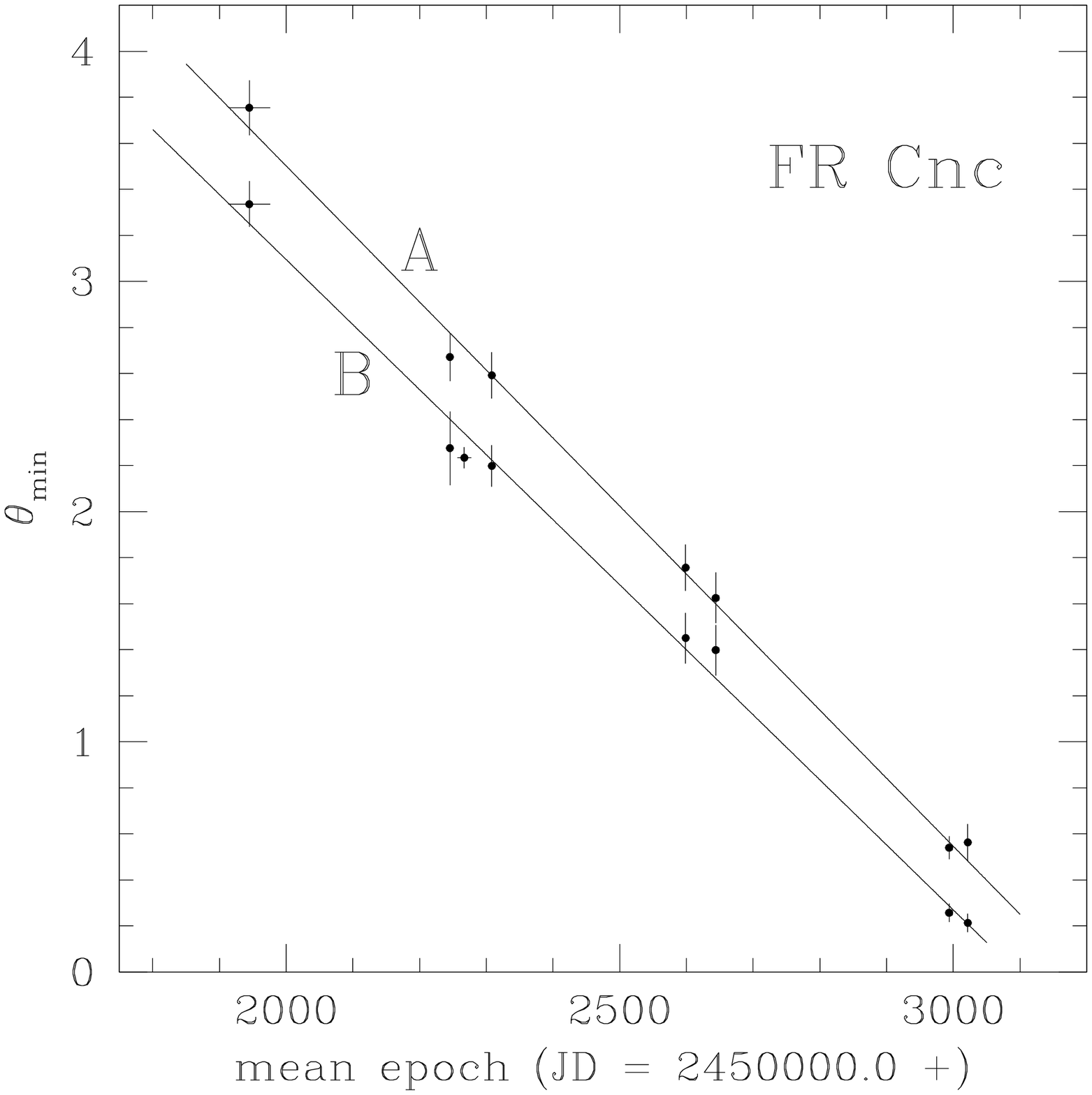}
\includegraphics[height=8cm,width=8cm]{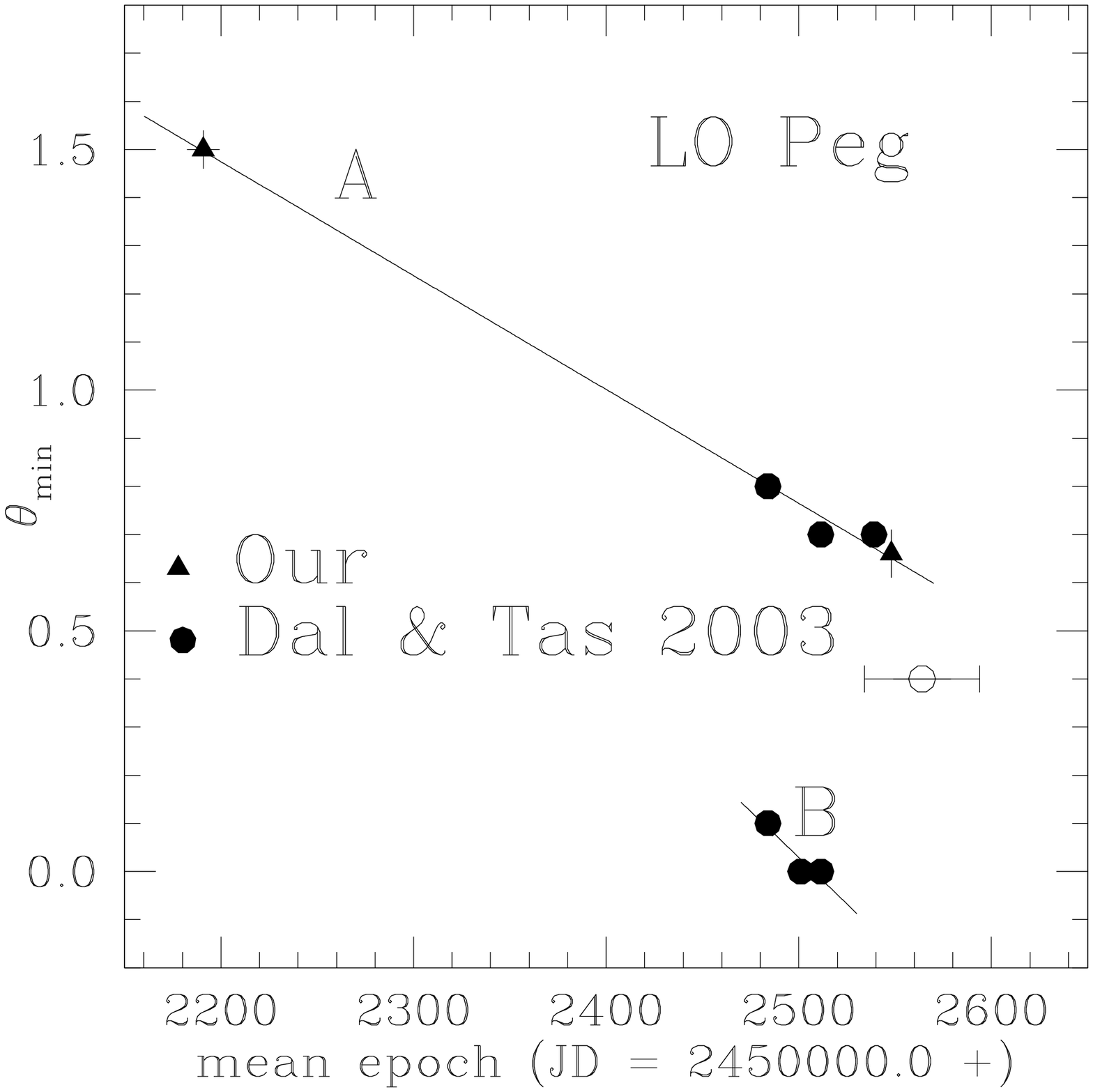}
}
\caption{ Plot of the mean epoch vs the phase minimum of light. {\it left panel:} 
for FR Cnc, {\it right panel:} for LO Peg, the open circle in the right panel 
was not used in the fitting.  Vertical bar shows the associated error in the 
determination of phase minimum and horizontal bar shows the length of epoch.}
\end{figure}

\clearpage
\begin{figure}
\hspace{-0.5cm}
\hbox{
\includegraphics[height=12cm,width=6.5cm]{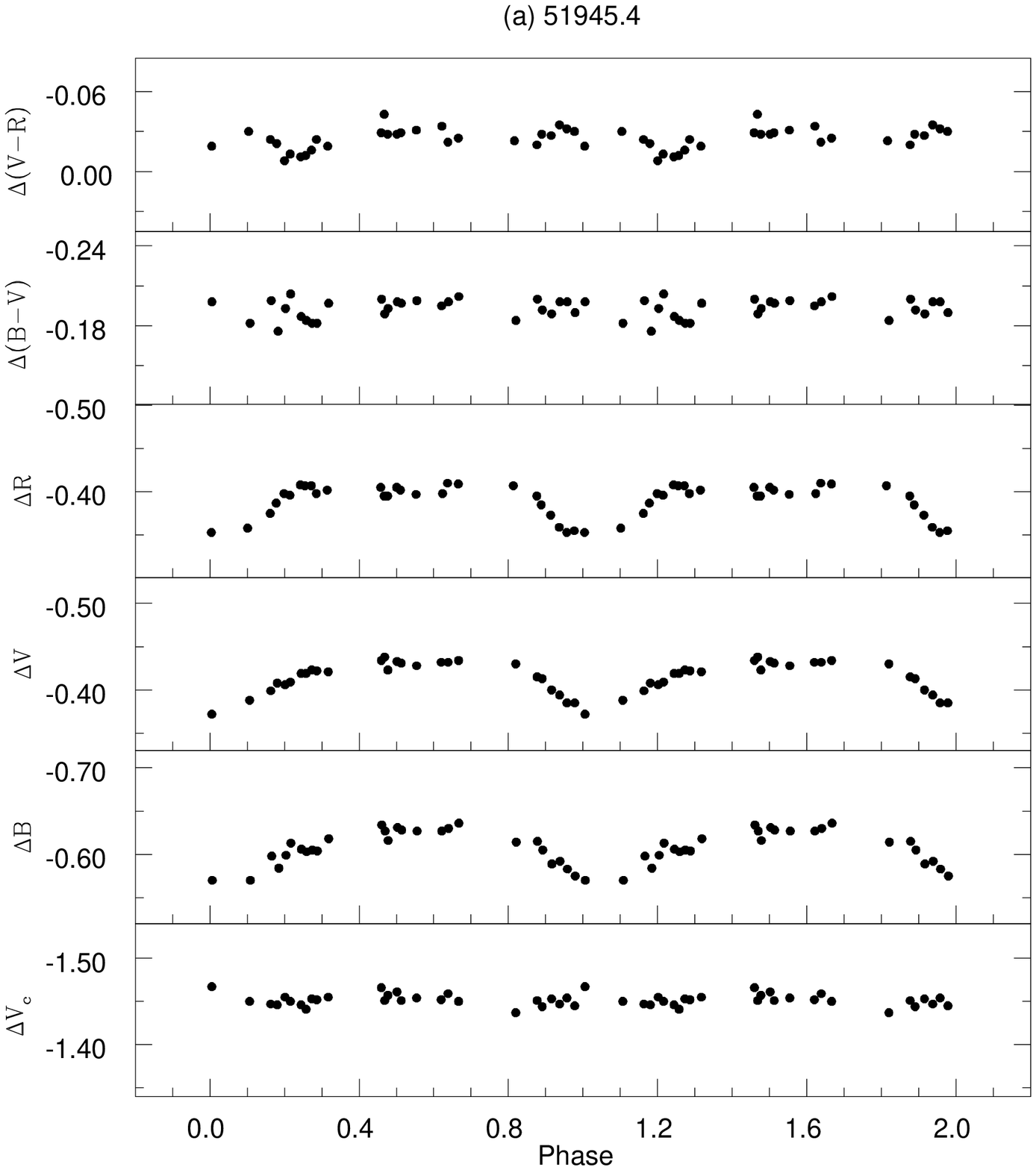}
\includegraphics[height=12cm,width=6.5cm]{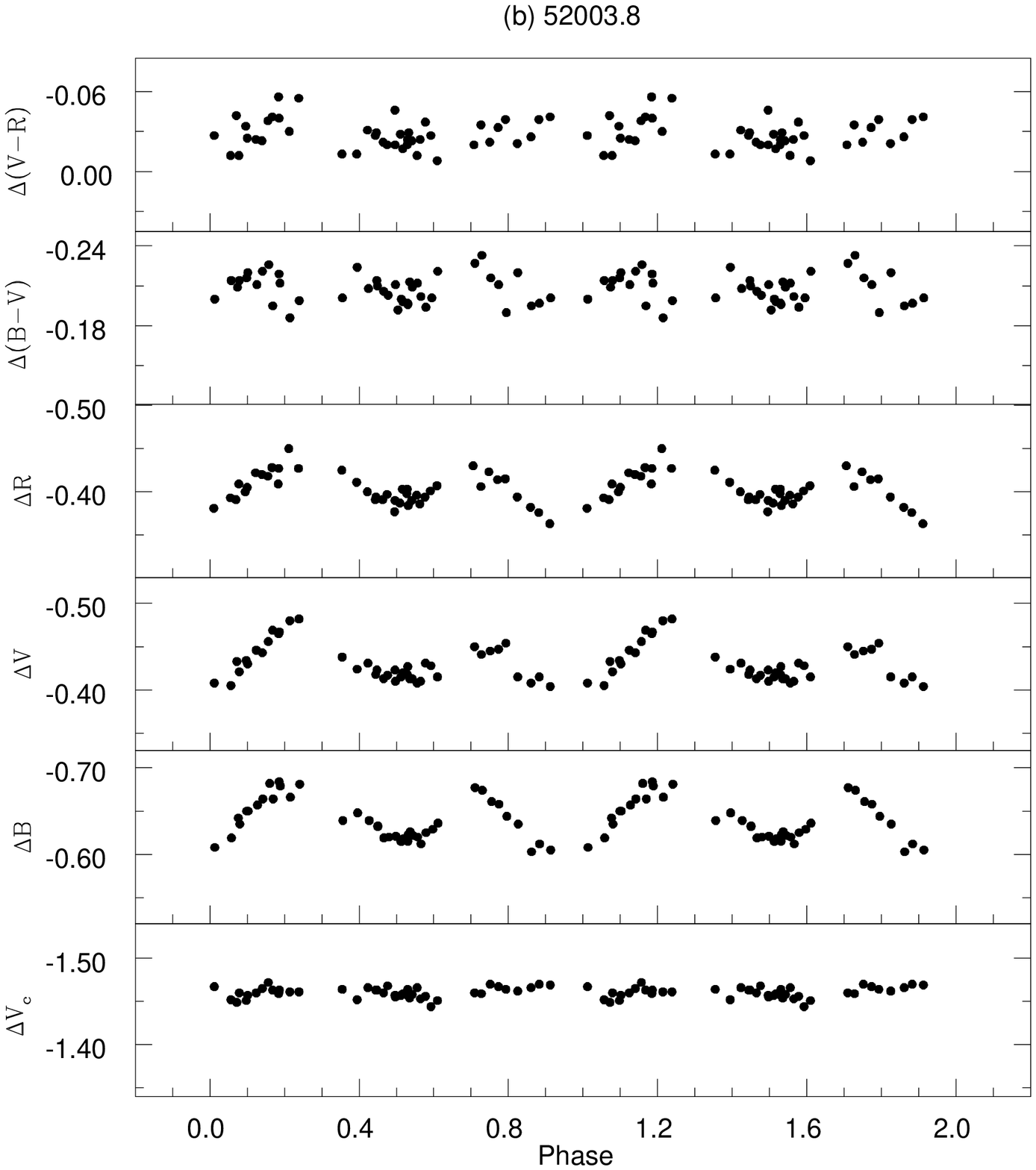}
}
\caption{$V_{c}, B, V, R $ light curves and (B-V), (V-R) color curves of
HD 95559 folded using a period of $1.52599775 \rm~{d}$, and shown for 
two different epochs.  The epoch (JD=2400000.0+) is marked at the top of each 
panel. Bottom panel in each Figure represents the plot of  different measures 
between the comparison (HD 95467) and the check (BD +33\degr 2294) star.}
\end{figure}
\clearpage

\begin{figure}
\hspace{-0.5cm}
\hbox{
\includegraphics[height=12cm,width=6.5cm]{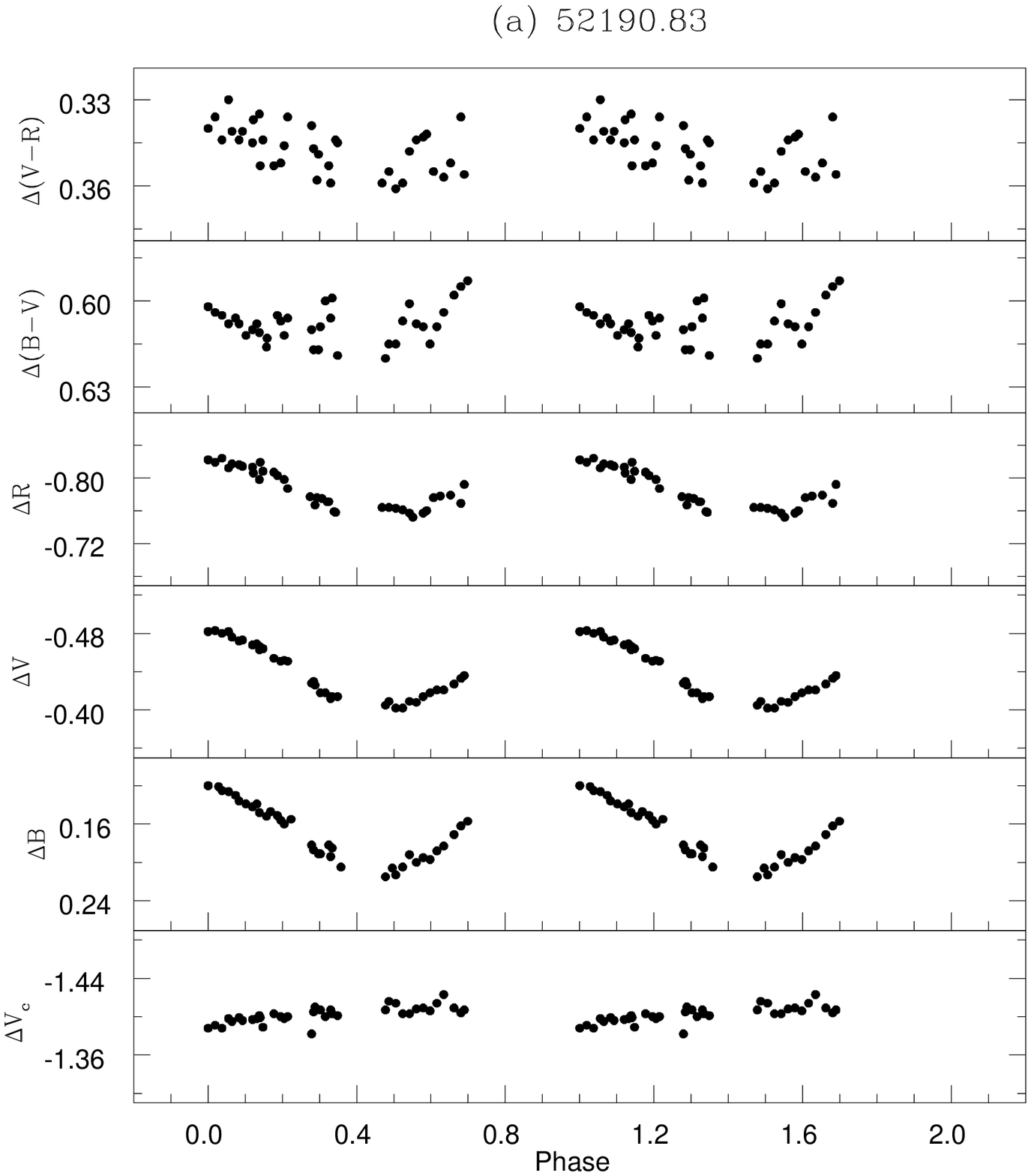}
\includegraphics[height=12cm,width=6.5cm]{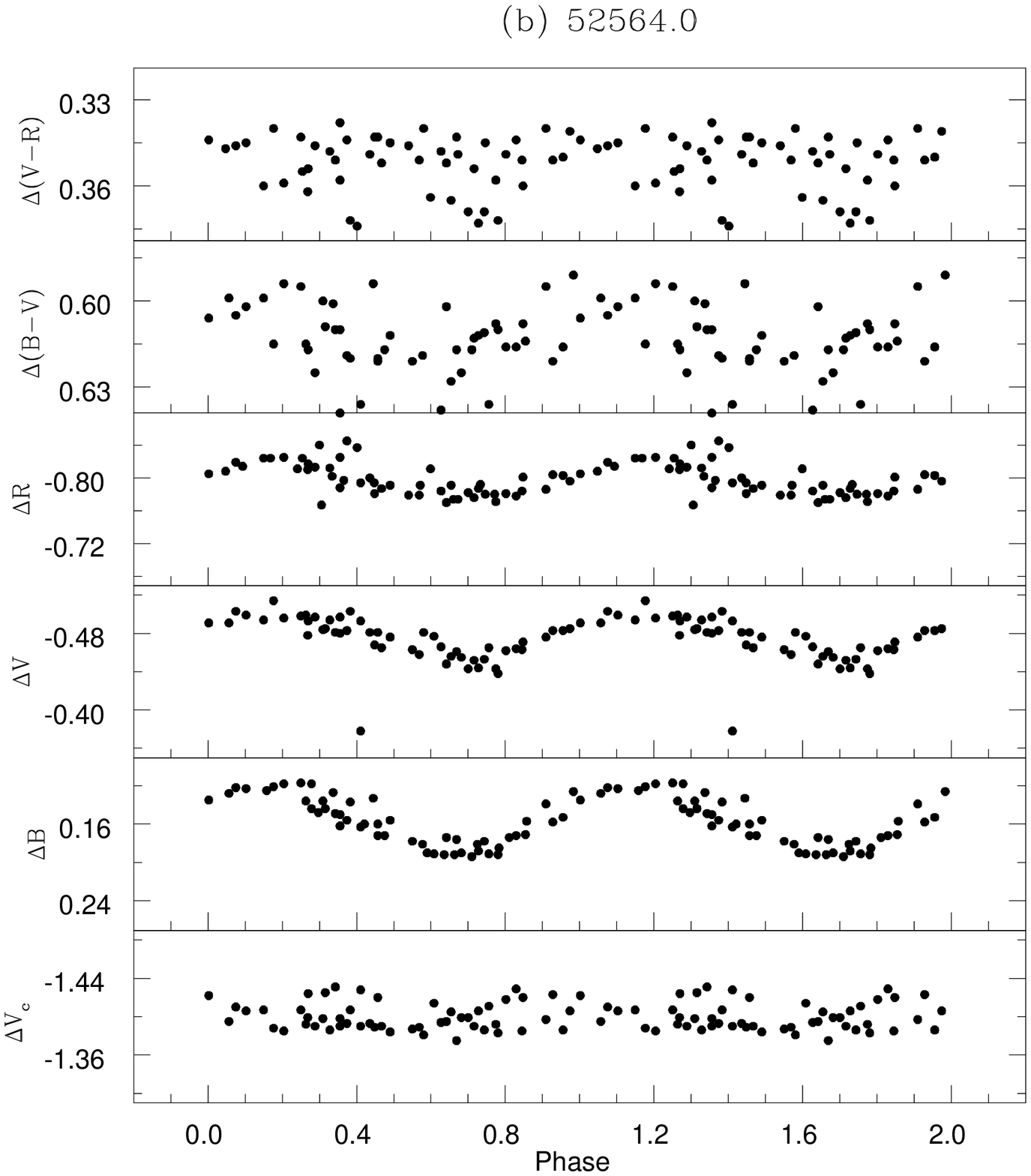}
}
\caption{$V_{c}, B, V, R $ light curves and (B-V), (V-R) color curves of
LO Peg folded using a period of $0.42375 \rm~{d}$, and shown for two 
different epochs.  The epoch  marked at the top of each panel is JD = 2400000.0+
. The bottom panel in each Figure represents the plot of different measures
of the comparison (BD +22\degr 4405) and the check (USNO-B1.0 1133-0542608) 
star.}
\end{figure}
\clearpage

\begin{figure}
\centering
\hspace{-1.0cm}
\includegraphics[height=16cm,width=10cm]{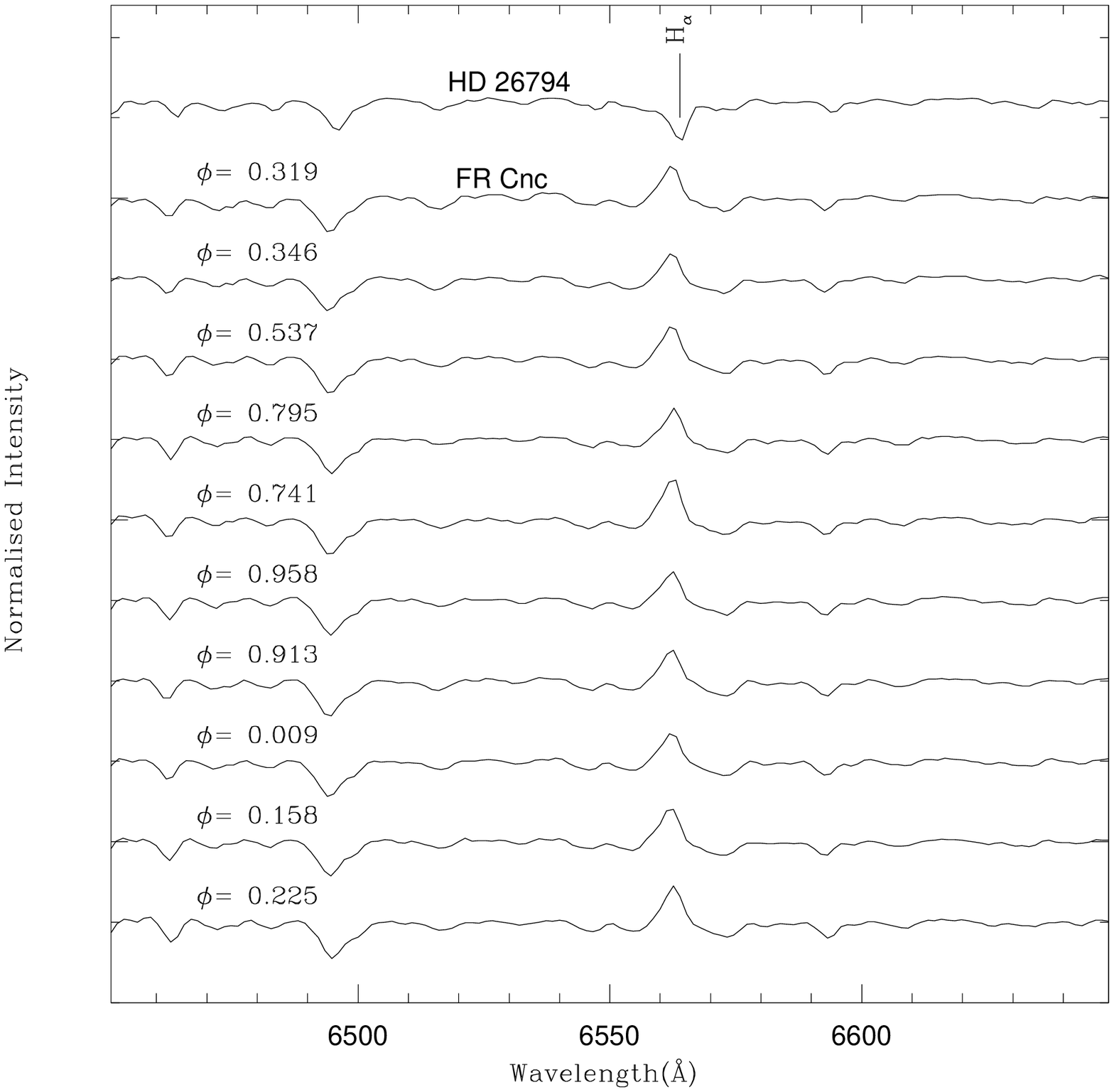}
\vspace{-0.5cm}
\caption{$H_{\alpha}$ spectra of the FR Cnc and the comparison star HD 26794.
The phase is mentioned at the top of each spectrum}.
\end{figure}

\clearpage

\begin{figure}
\centering
\includegraphics[height=14.0cm,width=10cm]{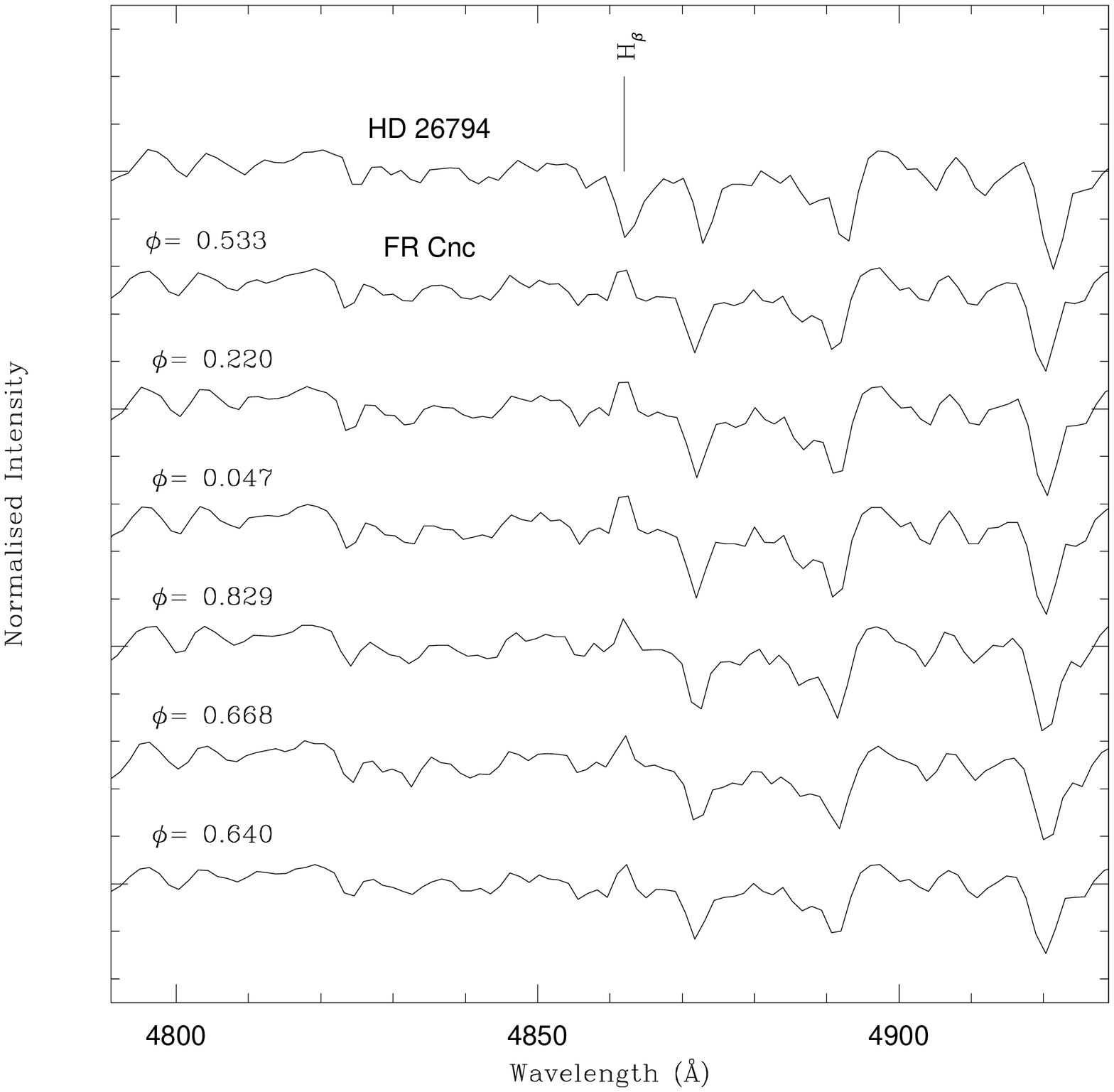}
\vspace{-0.5cm}
\caption{Spectra of FR Cnc showing $H_{\beta}$ in emission, while spectrum of 
HD 26794 shows $H_{\beta}$ in absorption.  The phase is mentioned at the top 
of each spectrum of FR Cnc.} 
\end{figure}

\clearpage
\begin{figure}[t]
\centering
\includegraphics[height=14cm,width=10cm]{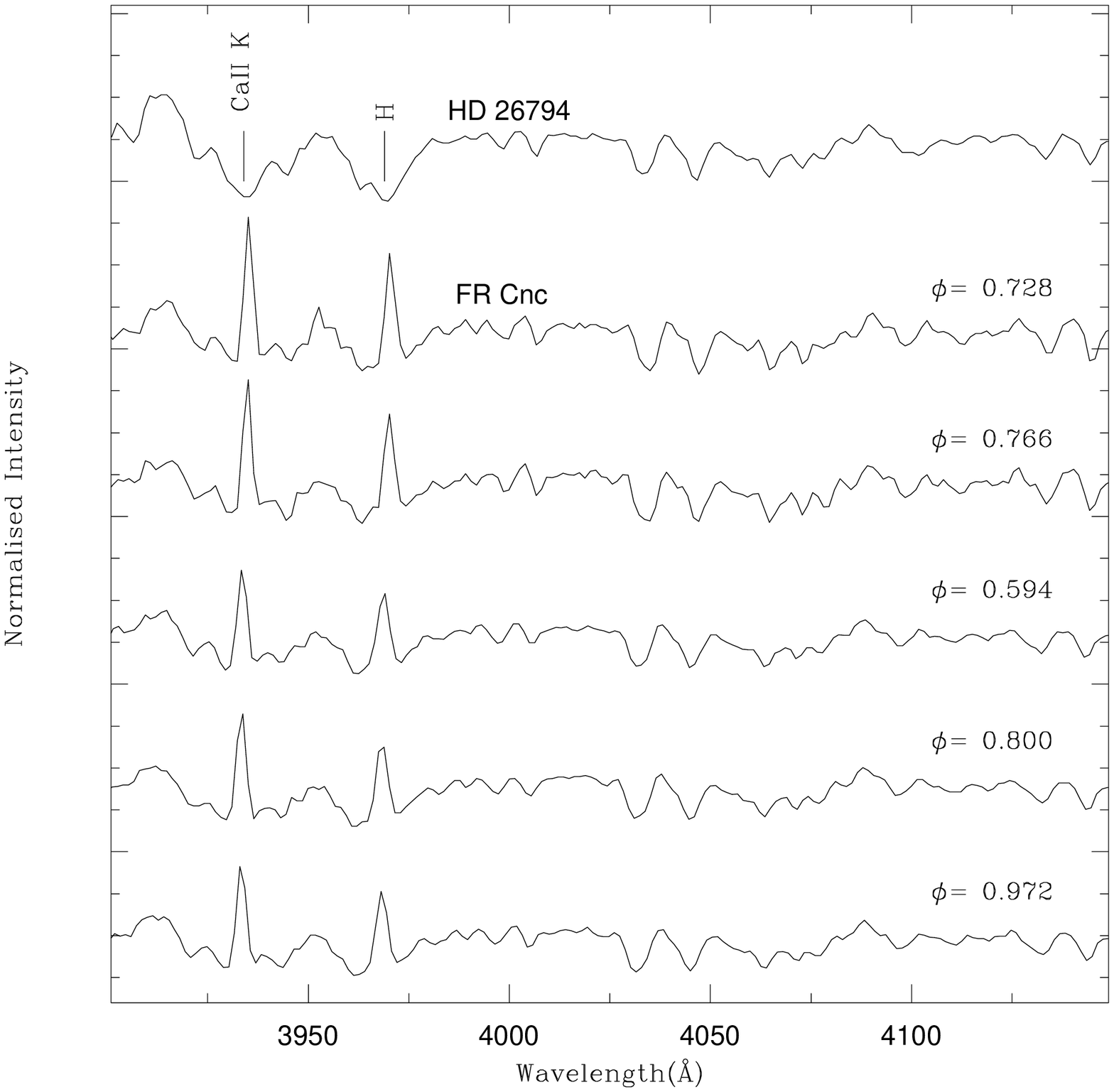}
\caption{Spectra of FR Cnc and the comparison star HD 26794 near the 
Ca II region. The phase is mentioned at the top corner of each spectrum of FR 
Cnc.}
\end{figure}
\clearpage
\begin{figure}[t]
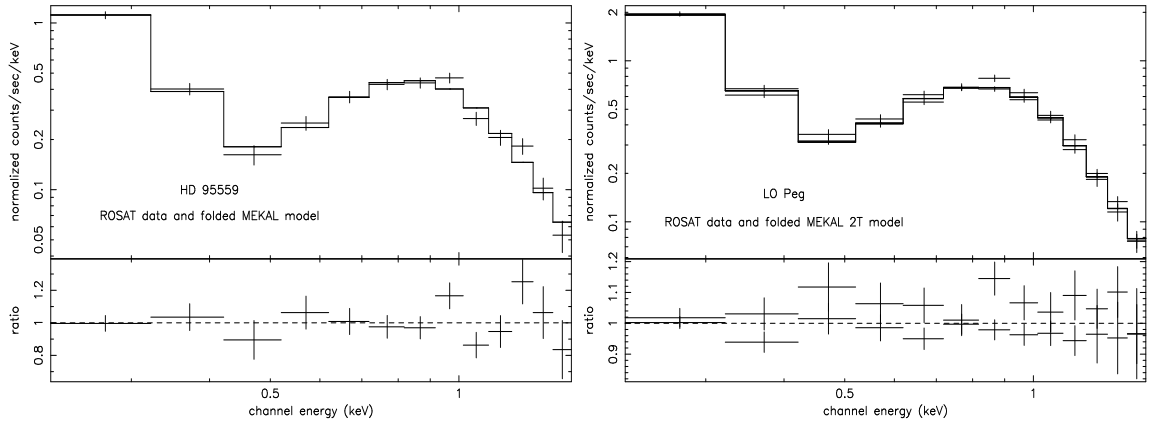

\label{xspec.fig}
\includegraphics[height=7.5cm,width=5.5cm,angle=-90]{f10a.eps}
\includegraphics[height=7.5cm,width=5.5cm,angle=-90]{f10b.eps}
\caption{Spectrum of HD 95559 (left panel) and LO Peg (right panel) with
the ROSAT PSPC detector, along with 2T MEKAL model fit and contribution to
ratio in each bin.}
\end{figure}
\clearpage
\begin{figure}
\includegraphics[height=12cm, width=12cm]{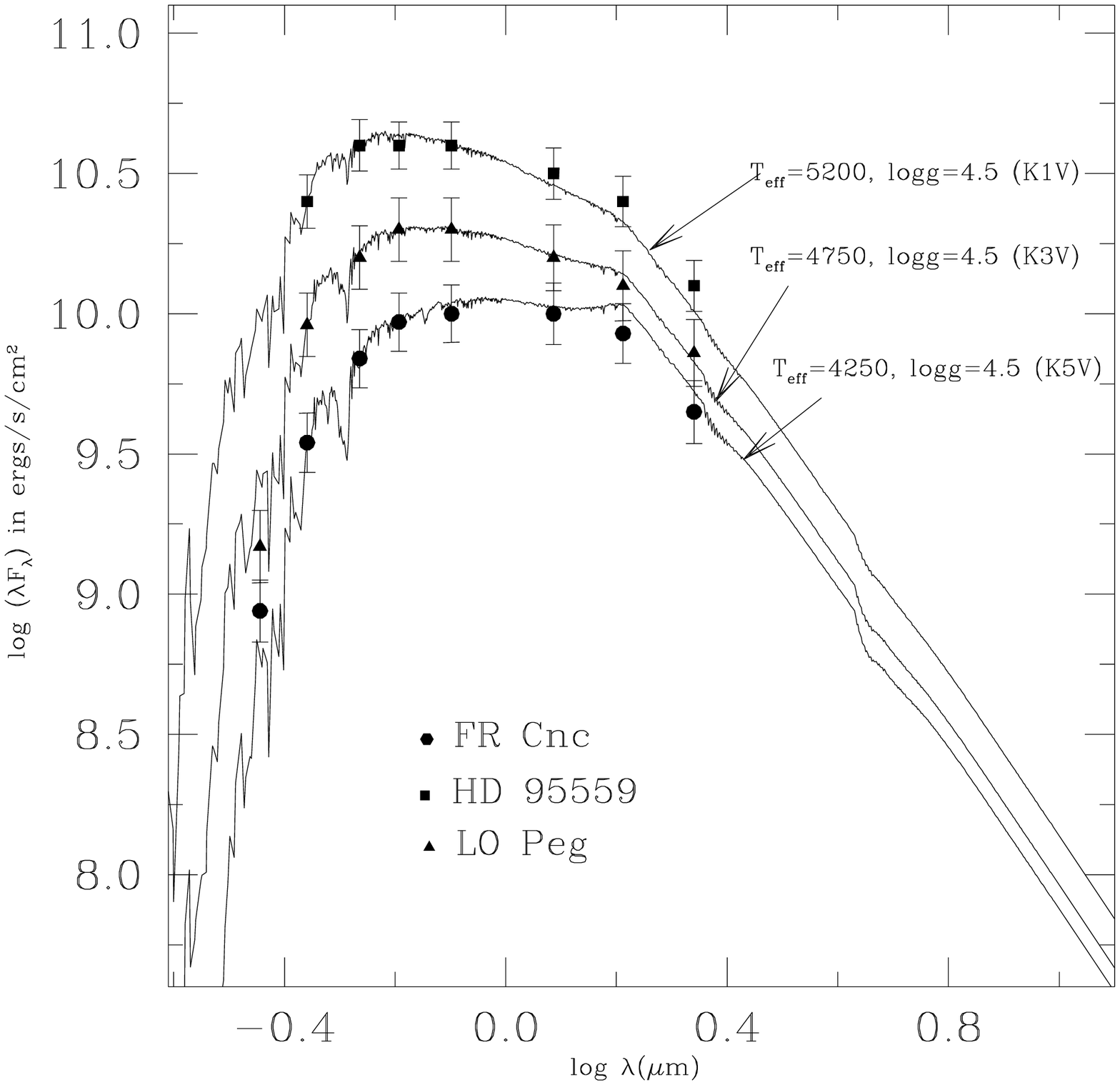}
\caption{SED of the stars FR Cnc (solid dots), HD 95559 (solid squares) and LO Peg (solid 
triangles). The solid lines represents the model SEDs from Kurucz (1993) as 
expected from the intrinsic properties of the star.  The vertical bars show
the uncertainty associated due to the distance and the radius of the star.}
\end{figure}
\end{document}